\documentclass[aip,jcp,reprint]{revtex4-1}
\usepackage{amsmath}
\usepackage{amsfonts}
\usepackage{amssymb}
\usepackage{graphicx}
\usepackage{color}
\usepackage{hyperref}

\newcommand \ket[1] {|{#1}\rangle}
\newcommand \bra[1] {\langle {#1}|}

\newcommand \trace[1] {\mathrm{Tr}({#1})}

\begin{document}

\title{Benchmarking density functional approximations for diamagnetic and paramagnetic molecules in nonuniform magnetic fields}

\author{Sangita Sen}

\email{sangita.sen310187@gmail.com}
\affiliation{
Department of Chemical Sciences, Indian Institute of Science, Education and Research, Kolkata, India}

\author{Erik I. Tellgren}

\email{erik.tellgren@kjemi.uio.no}
\affiliation{
Hylleraas Centre for Quantum Molecular Sciences, Department of Chemistry, University of Oslo, P.O.~Box 1033 Blindern, N-0315 Oslo, Norway}

\begin{abstract}
In this paper, correlated studies on a test set of 36 small molecules are carried out with both wavefunction (HF, MP2, CCSD) and density functional (LDA, KT3, cTPSS, cM06-L) methods.
The effect of correlation on exotic response properties such as molecular electronic anapole susceptibilities is studied and the performance of the various density functional approximations are benchmarked against CCSD and/or MP2.
Atoms and molecules are traditionally classified into `diamagnetic' and `paramagnetic' based on their isotropic response to uniform magnetic fields.
However, in this paper we propose a more fine-grained classification of molecular systems on the basis of their response to generally non-uniform magnetic fields. The relation of orientation to different qualitative responses is also considered.

\end{abstract}

\maketitle 

\section{Introduction} \label{intro}

Magnetic field effects pose unique challenges for quantum chemistry. Although the calculation of particular properties, most notably magnetic dipoles, nuclear shielding constants, and current densities induced by uniform magnetic fields, is nowadays routine, many other aspects have been subject to relatively few systematic studies, if any at all. Higher-order magnetic response and response to nonuniform magnetic fields are examples of this~\cite{Faglioni2004,Pagola2014,Caputo1994,Caputo1994a}. Moreover, nonperturbative effects of strong magnetic fields alter the normal chemistry of small molecules, giving rise to an exotic and largely unexplored strong field chemistry~\cite{TURBINER_PR424_309,Lange2012,Tellgren2012}. These challenges appear at all levels of theory, as for example time-reversal and spin symmetry and other built in adaptations to zero field settings need to be reconsidered. In density-functional theory, in particular, an additional aspect is that magnetic field effects are formally beyond the scope of the standard mathematical formulation, necessitating extensions~\cite{GRAYCE_PRA50_3089,VIGNALE_PRL59_2360,REIMANN_JCTC13_4089,TELLGREN_PRA97_012504}. Yet, the practically available density functional methods have almost exclusively been developed as heuristic approximations with the standard formulation, and approximations that properly incorporate magnetic fields are not yet mature enough to be practically useful~\cite{Tellgren2014}. Insofar as available density functional methods produce useful estimates of magnetic field effects, there is a high risk this is due to error cancellations specific to the most common magnetic properties. For example, some functionals have been fitted to magnetizabilities or nuclear shielding constants~\cite{Keal2004}. In the present work, we explore properties related to magnetic field gradients using several available density functional methods, including meta-GGAs that have recently emerged as particularly promising candidates for magnetic field effects, and compare to results at the second-order M\o{}ller--Plesset (MP2) and coupled-cluster singles and doubles levels (CCSD).

Quantum chemical computations of magnetic properties almost always rely on the assumption of weak-fields enabling formulations based on perturbation theory. 
However, for higher-order magnetic response, in particular when London atomic orbitals (LAOs) are employed to enforce gauge-origin invariance and accelerate basis set convergence~\cite{London1937,Hameka1958,Ditchfield1976,Helgaker1991}, the perturbative approach becomes increasingly more difficult. Ordinary Gaussian basis sets require very large basis sets for gauge-origin invariance~\cite{Faglioni2004,Caputo1994,Caputo1994a,Caputo1996,Caputo1997}. In this study we have used LAOs in combination with a non-perturbative (finite field) approach. Integral evaluation for the LAOs which are plane-wave/Gaussian hybrid functions~\cite{Tellgren2008,REYNOLDS_PCCP17_14280} has been implemented in the {\sc London} program~\cite{Tellgren2008,LondonProgram} and has been employed in the finite field computation of magnetic properties~\cite{Tellgren2008,Tellgren2012,Tellgren2013,Tellgren2014,Lutnaes2009,Sen2018a} earlier.
Since the introduction of the magnetic field in the Hamiltonian only requires a modification of the one-electron part, no additional implementation is necessary for extension to post-Hartree--Fock methods.
A non-perturbative approach also opens up the possibility of studying strong magnetic fields competing with the Coulomb forces and has led to the discovery of non-perturbative transition from closed-shell para- to diamagnetism~\cite{Tellgren2009} and a new bonding mechanism~\cite{Lange2012,Tellgren2012,Kubo2007} in very strong magnetic fields.

In this work we study the effect of correlation on anapole moments arising from induced orbital currents for a set of 36 closed shell molecules subject to transverse magnetic field gradients.
We have benchmarked LDA (SVN5)~\cite{VOSKO_CJP58_1200}, KT3~\cite{Keal2004}, cTPSS~\cite{TAO_PRL91_146401} and cM06-L~\cite{ZHAO_JCP125_194101,ZHAO_JPCA112_6794} functionals against CCSD and/or MP2.
An earlier study~\cite{ZARYCZ_JCC37_1552} to assess the effects of correlation was inconclusive. CCSD results were reported in aug-cc-pVDZ basis which is too small for accurate representation of such high order properties. Moreover, the relative quality of the density functional approximations studied (KT3~\cite{Keal2004}, B3LYP~\cite{BECKE_JCP98_5648}, CAMB3LYP~\cite{YANAI_CPL393_51}) could not be established. Basis set convergence of the anapole susceptibility values was well-studied and MODENA basis sets were proposed to be superior to Dunning's basis sets.
Previous studies of density functional approximations for magnetic properties like magnetizabilities and NMR shielding constants~\cite{Tellgren2014,Furness2015,Reimann2019} have indicated that meta-GGA functionals, in particular cTPSS, are promising candidates for capturing even exotic magnetic effects very far from the domain these functionals were explicitly constructed for. Remarkably, it was also found that the errors for the paramagnetic closed shell molecules were an order of magnitude higher for all the methods studied except for cTPSS (current TPSS).
 The present work investigates whether this trend holds for more exotic magnetic properties like anapole susceptibilities, which have not been considered thus far.

Because paramagnetic systems typically exhibit stronger correlation effects than diamagnetic systems, we will below generalize these concepts to allow for nonuniform magnetic fields.
Historically, the discovery of diamagnetism is credited to Anton Brugmans who observed in 1778 that bismuth was repelled by magnetic fields~\cite{Ben-Menahem2009}. William Whewell suggested the terminology \emph{diamagnetic} for materials repelled by a magnetic field and \emph{paramagnetic} for those attracted by it and Faraday adopted this~\cite{Jackson2015}.
The quantum picture of atomic and molecular magnetism was established with van Vleck's theory of paramagnetism and crystal field theory for solid-state magnetism, Dorfman's corresponding theory for metals, Pauli's work including the derivation of temperature independent paramagnetism, and Landau's quantum theory of diamagnetism.
Here we have bypassed the vast field of ferromagnetism, since this paper is not concerned with it. 

In modern terms, dia- and paramagnetism are understood in terms of whether the (second-order) response of the energy to a uniform field is positive or negative. In the mathematical literature, this distinction has also been used for arbitrary, nonuniform magnetic fields~\cite{Simon1976,Erdos1997}. Because a uniform field is a three-dimensional vector, the second-order magnetic susceptibility is a $3\times 3$ tensor. Taking into account a sign convention, clear-cut examples of diamagnetism (or paramagnetism) occur when this tensor only has negative (or positive) eigenvalues. However, the tensor may also have both positive and negative eigenvalues, corresponding to decreasing energy for some, but not all, magnetic field components.
Conventionally, the sign of its isotropic value decides the classification of the molecule as dia- or paramagnetic. For example, the BH molecule is found to have a diamagnetic response to fields parallel to the chemical bond and a paramagnetic response to fields perpendicular to it. In this case, the isotropic magnetizability turns out to be paramagnetic as well leading to the classification of BH as a closed shell paramagnetic molecule. However, a molecule such as square C$_4$H$_4$ shows a weak paramagnetic response to fields perpendicular to the molecular plane but an overall diamagnetic response. 

The magnetic susceptibilities related to inhomogeneities in the magnetic field such as the anapole susceptibilities are independent of the magnetizabilities and may in some cases oppose the effects of the uniform component of the magnetic field. In our opinion, a classification should also encompass molecular response to non-uniform magnetic fields in general, as far as possible. In what follows, we propose a simple classification of magnetic response to nonuniform fields.
The response of the electrons to inhomogeneities in the external magnetic field arise from both orbital effects and spin effects. Among the few studies of these effects is the work by Lazzeretti and co-workers on a perturbative formalism for the orbital response due to field gradients~\cite{Lazzeretti1989,Lazzeretti1993} and some other studies at the H\"uckel-level~\cite{Ceulemans1998}, Hartree--Fock level~\cite{Faglioni2004,Caputo1994,Caputo1994a,Caputo1996,Caputo1997}, and correlated levels~\cite{ZARYCZ_JCC37_1552,SANDRATSKII_JPCM4_6927}.
While spin effects are certainly important and in most cases the dominant effect~\cite{Sen2018a,Sun2019}, this paper is only concerned with orbital effects in closed-shell molecules. Further exploratory studies are planned but beyond the present scope.

The response to (transverse) magnetic field gradients may be quantified by the \textit{anapole moments}~\cite{Zeldovich1957} which couple linearly to the curl of the magnetic field~\cite{Gray2010,Marinov2007,Spaldin2008,Kaelberer2010,Ogut2012,Ye2013}.
They may be considered to arise from the meridional currents in a toroidal charge distribution.
They are anti-symmetric under both spatial inversion and time-reversal.
Nuclear anapole moments are studied by physicists~\cite{Haxton1997,Haxton2002} in connection with parity violation with the first experimental evidence coming from measurements on the Cs atom~\cite{Wood1997,Haxton2001,DeMille2008}.
Experiments for measuring permanent and induced electronic anapole moments have been suggested\cite{Spaldin2008,Pelloni2011,Khriplovich1990}.
However, only special structures such as molecular nanotoroids~\cite{Ceulemans1998,Berger2012}, ferroelectric nanostructures~\cite{Naumov2004,VanAken2007}, ferromagnetic structures~\cite{Klui2003} and Dy clusters (single molecule magnets)~\cite{Novitchi2012,Guo2012,Ungur2012} are expected to have permanent anapole moments.
Anapole moments in metamaterials have also been observed with potential application in sensors~\cite{Kaelberer2010,Ye2013,Basharin2015a}.
On the other hand, induced anapole moments easily arise in molecules placed in external non-uniform fields and we can compute the corresponding susceptibilities. Both toroidal spin and/or orbital currents can give rise to anapole moments.
Induced anapolar current densities in conjugated cyclic acetylenes~\cite{Berger2012} and some small molecules~\cite{Pelloni2011} have been studied.
Spin and orbital contributions to anapole moments have been analyzed in a simple analytical model of diatomics\cite{Khriplovich1990,Lewis1994} and also using non-perturbative General Hartree--Fock theory~\cite{Sen2018a}.
The orbital contributions have been estimated by both perturbative approaches~\cite{Faglioni2004,Caputo1994,Caputo1994a,Caputo1996,Caputo1997} and non-perturbative approaches~\cite{Tellgren2013}.
Faglioni et al\cite{Faglioni2004} have derived the perturbative expressions for induced anapole moments.

The outline of the article is as follows.
In Sec.~\ref{definitions}, we define the Hamiltonian and the properties relevant to our study.
Sec.~\ref{classification} discusses our proposed classification of molecules.
Sec.~\ref{results} presents our results on the effect of correlation on the anapole susceptibilities and the relative performance of the various density functional approximations. Finally, we conclude with the summary in Sec.~\ref{summary}.

\section{Hamiltonian and Properties} \label{definitions}

In this study, the non-uniform magnetic field has the form,
\begin{equation} \label{magfield}
  \mathbf{B}_{\text{tot}}(\mathbf{r}) = \mathbf{B} + \mathbf{r}_{\mathbf{h}}^T \mathbf{b} - \frac{1}{3}\mathbf{r}_{\mathbf{h}} \, \trace{\mathbf{b}},
\end{equation}
where $\mathbf{B}$ is a uniform (position independent) component, $\mathbf{b}$ is a $3\times 3$ matrix defining the field gradients, and $\mathbf{r}_{\mathbf{h}} = \mathbf{r} - \mathbf{h}$ is the position relative to some reference point $\mathbf{h}$. 
This form may be viewed as arising from a truncation of a Taylor expansion of a general magnetic field around $\mathbf{r} = \mathbf{h}$ at linear order.
The corresponding vector potential can be written as
\begin{align}
\mathbf{A}_{\text{tot}}(\mathbf{r}) &= \frac{1}{2}\mathbf{B} \times \mathbf{r_g}-\frac{1}{3}\mathbf{r_h} \times (\mathbf{r_h}^T\mathbf{b}),
\end{align}
where $\mathbf{r}_{\mathbf{g}} = \mathbf{r} - \mathbf{g}$, $\mathbf{g}$ being the gauge origin. One can show that $\mathbf{B}_{\text{tot}} = \nabla\times\mathbf{A}_{\text{tot}}$ and that the magnetic field is divergence free, $\nabla\cdot\mathbf{B}_{\text{tot}} = 0$. 
The symmetric part, $\mathbf{b} = \mathbf{b}^T$, can be set to zero and
we focus on the anti-symmetric part C$_{\alpha} = \epsilon_{\alpha\beta\gamma} b_{\beta\gamma}$ of the matrix $\mathbf{b}$. 
We can then write
\begin{align}
  \mathbf{A}_{\text{tot}}(\mathbf{r}) & = \frac{1}{2}\mathbf{B} \times \mathbf{r}_{\mathbf{g}}-\frac{1}{3} \mathbf{r}_{\mathbf{h}} \times (\mathbf{C}\times\mathbf{r}_{\mathbf{h}}),
         \\
  \mathbf{B}_{\text{tot}}(\mathbf{r}) & = \mathbf{B} + \frac{1}{2} \mathbf{C}\times\mathbf{r}_{\mathbf{h}}.
\end{align}
Furthermore, the anti-symmetric part of $\mathbf{b}$  equals the curl of the magnetic field, $\nabla \times \mathbf{B}_{\text{tot}} = \mathbf{C}$ which is taken to be position independent.

The non-relativistic Schr\"odinger--Pauli Hamiltonian is given by
\begin{equation}
\hat{H} = \frac{1}{2}\sum_l \hat{\pi_l}^2 - \sum_l v(\mathbf{r}_l) + \sum_{k<l} \frac{1}{r_{kl}} + \sum_l\mathbf{B}_{\text{tot}}(\mathbf{r}_l)\cdot \hat{\mathbf{S}}_l \label{Hamiltonian}
\end{equation}
where $\hat{\boldsymbol{\pi}}_l = -i\nabla_l + \mathbf{A}_{\text{tot}}(\mathbf{r}_l)$ is the mechanical momentum operator. Properties can be alternately viewed as expectation values $\bra{\Psi} \hat{\Omega} \ket{\Psi}$ or as derivatives of the energy $E = \bra{\Psi} \hat{H} \ket{\Psi}$ related to terms in a Taylor expansion. In this study we ignore the spin-Zeeman term and thereby the spin-breaking induced by the non-uniform part of the magnetic field.
The first order orbital angular moment,
\begin{align}
   \mathbf{L}_{\mathbf{q}} & = \sum_l \bra{\Psi} \hat{\mathbf{L}}_{\mathbf{q};l}\ket{\Psi}, \quad \hat{\mathbf{L}}_{\mathbf{q};l} = \mathbf{r}_{\mathbf{q};l} \times \hat{\boldsymbol{\pi}}_l,
\end{align}
is with respect to an arbitrary reference point, $\mathbf{q}$. Given the form of the magnetic vector potential above, it is $\mathbf{L}_{\mathbf{g}}$, with the reference point at the gauge origin, that is the relevant magnetic dipole moment. The orbital anapole moment is similarly given by,
\begin{equation}
      \mathbf{a} = -\sum_l \bra{\Psi} \mathbf{r}_{\mathbf{h};l} \times \tfrac{1}{3} \hat{\mathbf{L}}_{\mathbf{h};l} \ket{\Psi}.
\end{equation}.

Recalling that the current density can be obtained as the functional derivative $\mathbf{j} = \delta E/\delta \mathbf{A}_{\mathrm{tot}}$, we can also identify he magnetic orbital dipole moment and anapole moment  with linear and quadratic moments of the current density,
\begin{align}
    \mathbf{J}_{\mathbf{g}} & = \int \mathbf{r}_{\mathbf{g}} \times \mathbf{j}(\mathbf{r}) \, d\mathbf{r},
       \\
    \mathbf{a} & = -\frac{1}{3} \int \mathbf{r}_{\mathbf{h}} \times \big( \mathbf{r}_{\mathbf{h}} \times \mathbf{j}(\mathbf{r}) \big) \, d\mathbf{r}.
\end{align}

We note that the energy $E$ as well as expectation value properties like $\mathbf{J}_{\mathbf{g}}$ and $\mathbf{a}$ can be obtained directly as functions of $\mathbf{B}$ and $\mathbf{C}$. We can thus,define second-order properties from a Taylor expansion of the energy
\begin{equation}
  \label{eqEnergyExpansion}
  \begin{split}
   E(\mathbf{B},\mathbf{C}) & \approx E_0 + \frac{1}{2}\mathbf{B} \cdot \mathbf{J}_{\mathbf{g}} - \frac{1}{2} \mathbf{C}\cdot \mathbf{a} -  \frac{1}{2} \mathbf{B}^T \boldsymbol{\chi} \mathbf{B} \\
      & \ \ \ \ - \mathbf{B} \mathcal{M} \mathbf{C} - \frac{1}{4} \mathbf{C}^T \mathcal{A} \mathbf{C},
  \end{split}
\end{equation}
where $\mathbf{J}_{\mathbf{g}}$ and $\mathbf{a}$ are evaluated at $\mathbf{B}_{\text{tot}}=0$. We can identify $\boldsymbol{\chi}$ as the magnetizability tensor, and call $\mathcal{M}$ as the mixed anapole susceptibility tensor, and $\mathcal{A}$ as the anapole susceptibility tensor.

When the Hellmann--Feynman theorem is applicable, the expectation value quantities can be equated with energy derivatives,
\begin{align}
   \mathbf{J}_{\mathbf{g}} & \stackrel{!}{=} 2 \frac{\partial E(\mathbf{B},\mathbf{C})}{\partial \mathbf{B}},
         \\
   \mathbf{a} & \stackrel{!}{=} -2 \frac{\partial E(\mathbf{B},\mathbf{C})}{\partial \mathbf{C}}.
\end{align}
However, when LAOs are used, the basis set depends on the parameters $\mathbf{B}$ and $\mathbf{C}$ leading to a discrepancy between the expectation values and the energy derivatives, in general, except in the complete basis set limit.

Second-order susceptibilities may be defined as follows:
\begin{align}
        \chi_{\alpha\beta} &= -\frac{\partial^2 E(\mathbf{B},\mathbf{C})}{\partial B_\alpha\partial B_\beta}\bigg|_{\mathbf{B}=0,\mathbf{C}=0}, \label{magn_def}\\
		\mathcal{A}_{\alpha\beta} &= -2\frac{\partial^2 E(\mathbf{B},\mathbf{C})}{\partial C_\alpha\partial C_\beta}\bigg|_{\mathbf{B}=0,\mathbf{C}=0}, \label{Asus_def}\\
		\mathcal{M}_{\alpha\beta} &= -\frac{\partial^2 E(\mathbf{B},\mathbf{C})}{\partial B_\alpha\partial C_\beta}\bigg|_{\mathbf{B}=0,\mathbf{C}=0}. \label{Msus_def}
\end{align}
One can also introduce the closely related, but inequivalent, quantities
\begin{align}
		\mathcal{A}_{\alpha\beta}' &= \frac{\partial a_\alpha(\mathbf{B},\mathbf{C})}{\partial C_\beta}\bigg|_{\mathbf{B}=0,\mathbf{C}=0}, \label{Aprimesus_def}\\
		\mathcal{M}_{\alpha\beta}' &= -\frac{1}{2}\frac{\partial L_{\mathbf{g};\alpha}(\mathbf{B},\mathbf{C})}{\partial C_\beta}\bigg|_{\mathbf{B}=0,\mathbf{C}=0},  \label{Mprimesus_def}\\
		\mathcal{M}_{\alpha\beta}'' &= \frac{1}{2}\frac{\partial a_{\beta}(\mathbf{B},\mathbf{C})}{\partial B_\alpha}\bigg|_{\mathbf{B}=0,\mathbf{C}=0}.  \label{Mdblprimesus_def}
\end{align}
Again, in the basis set limit, equivalence is restored, i.e., $\mathcal{A} = \mathcal{A}'$ and $\mathcal{M} = \mathcal{M}' = \mathcal{M}''$.
Note that the multiplicative factors in Eqs.~\ref{eqEnergyExpansion}, \ref{Msus_def}, \ref{Mprimesus_def} and \ref{Mdblprimesus_def} have been corrected from those reported in earlier publications~\cite{Tellgren2013,Sen2018a} in order to be self-consistent with the other definitions. This implies that the $\mathcal{M}$ values reported in these publications should be halved. However, this has no implication on the conclusions of the two papers.

\section{Results and Discussion} \label{results}

Our test set contains 36 molecules (HF, CO, N$_2$, H$_2$O, HCN, HOF, LiH, NH$_3$, H$_2$CO, CH$_4$, C$_2$H$_4$, AlF, CH$_3$F, C$_3$H$_4$, FCCH, FCN, H$_2$S, HCP, HFCO, H$_2$C$_2$O, LiF, N$_2$O, OCS, H$_4$C$_2$O, PN, SO$_2$, OF$_2$, H$_2$, H$_2$O$_2$, BH, CH$^+$, AlH, BeH$^-$, SiH$^+$, C$_4$H$_4$, FNO) and subsumes the test set of diamagnetic molecules in Tellgren et al.~\cite{Tellgren2014} and the closed shell paramagnetic molecules in the test set of Reimann et al.~\cite{Reimann2019}.
Geometries for the molecules are as reported in earlier publications~\cite{Tellgren2013,Tellgren2014,Reimann2019} and are also provided in the Supplementary Information.

All calculations were performed using the {\sc London} program~\cite{Tellgren2008,LondonProgram}. The density-functional calculations used the previously reported implementations~\cite{Tellgren2014,Furness2015}. The coupled-cluster calculations were performed using the previously reported implementation~\cite{Stopkowicz2015} as well as newer functionality~\cite{StopkowiczCCgrad}.
The symmetric finite difference formula for numerical second derivatives of the energy was employed to compute the anapole susceptibilities $\mathcal{A}$ and $\mathcal{M}$. Step sizes of $\epsilon =0.01$a.u. for $\mathbf{B}$ and $\epsilon'=0.005$~a.u. for $\mathbf{C}$ were used.
$\epsilon'$ was chosen to be smaller as the effect of $\mathbf{C}$ on the local magnetic field is scaled by the interatomic distances in the molecule.
The reference point, $\mathbf{h}$, for $\mathbf{C}$ was placed at the centre of charge of the molecules in all cases.
The error in the energy is quadratic in the step size within the limits to which the energy is converged while the error in the analytically computed moments (first derivative of energy) is linear.
All numerical results presented in this paper are given in SI-based atomic units---see earlier work for the conversion factors to SI units~\cite{Tellgren2013}.

The uncontracted aug-cc-pCVTZ basis set has been employed for all the computations. 
The name of the basis set is prefixed with `L' to denote the use of London atomic orbitals and `u' to indicate that the basis sets are uncontracted - Luaug-cc-pCVTZ. 

\subsection{Current-dependence and meta-GGAs}

The meta-GGA functional form allows a dependence on the kinetic energy density. In the absence of a magnetic field, the everywhere positive, canonical kinetic energy $\tau_{\mathrm{can}} = \tfrac{1}{2} \sum_k |\nabla\phi_k|^2$, with summation over occupied orbitals $\phi_k$, is the natural choice. In the presence of a magnetic field, $\tau_{\mathrm{can}}$ is gauge dependent and cannot be used. An obvious solution is to use the gauge independent, physical kinetic energy density $\tau_{\mathrm{phys}} = \tfrac{1}{2} \sum_k |(-i\nabla+\mathbf{A}_{\mathrm{tot}})\phi_k|^2$ instead. This choice has been suggested by Maximoff and Scuseria~\cite{MAXIMOFF_CPL390_408}. An alternative, with some theoretical aspects and numerical results in its favour~\cite{BECKE_CJC74_995,BATES_JCP137_164105,Reimann2019,Sen2018b}, is to instead use Dobson's gauge-invariant kinetic energy density,
\begin{equation}
    \tau_D = \tau_{\mathrm{phys}} - \frac{j^2}{2\rho} = \tau_{\mathrm{can}} - \frac{j_p^2}{2\rho},
\end{equation}
where $\mathbf{j}_p = \mathrm{Im} \sum_k \phi_k^* \nabla \phi_k$ is the paramagnetic current density and $\mathbf{j} = \mathrm{Re} \sum_k \phi_k^* (-i\nabla+\mathbf{A}_{\mathrm{tot}}) \phi_k$ is the physical current density.

Previous work~\cite{Reimann2019} used a prefix `a' to denote meta-GGA functionals using the physical $\tau_{\mathrm{phys}}$ (e.g., aTPSS) and a prefix `c' to denote functionals using Dobson's $\tau_D$ (e.g., cTPSS). In the present work we only consider the latter type, specifically cTPSS and cM06-L.

\subsection{Classification} \label{classification}

\begin{figure}
    \centering
    \includegraphics[width=\linewidth]{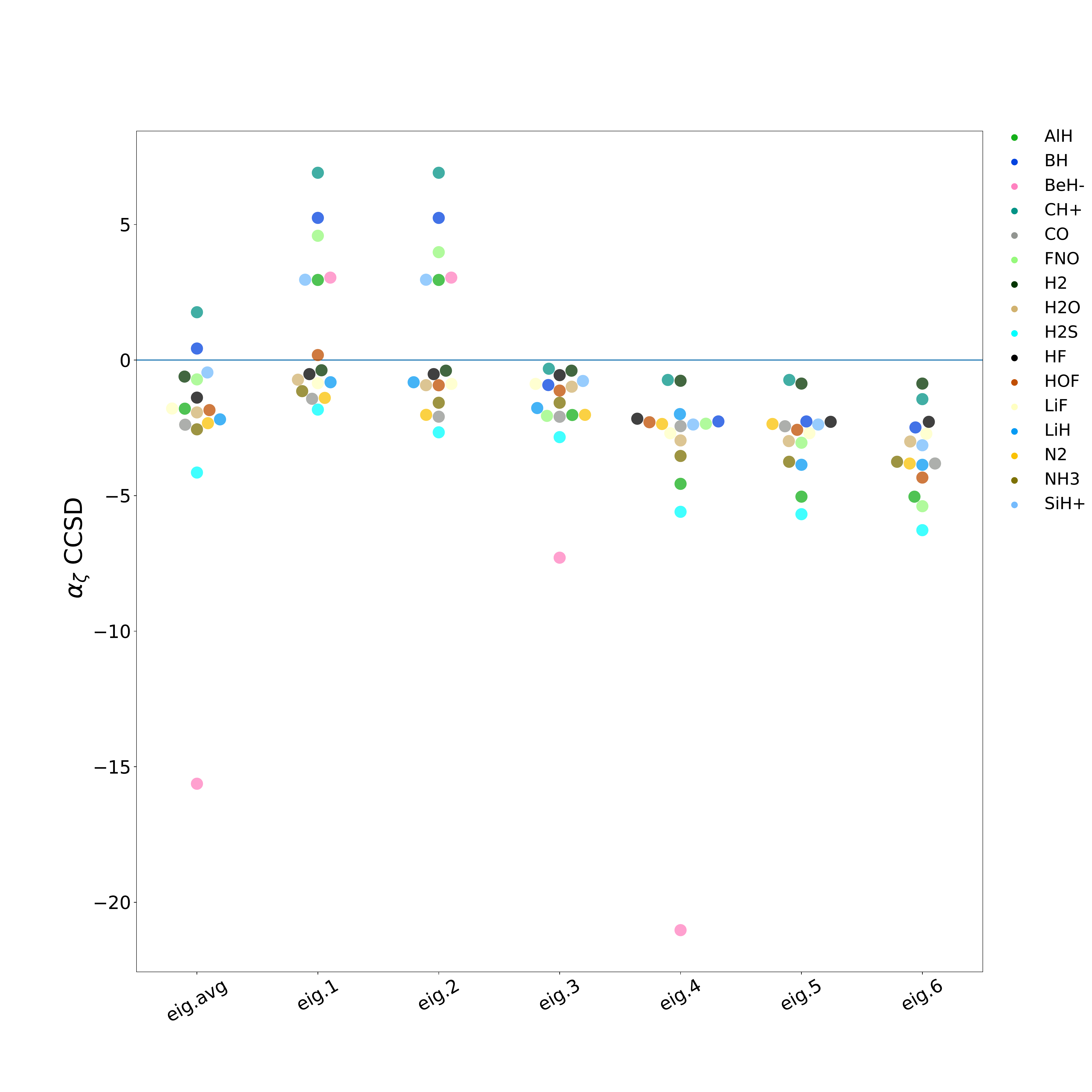}
    \includegraphics[width=\linewidth]{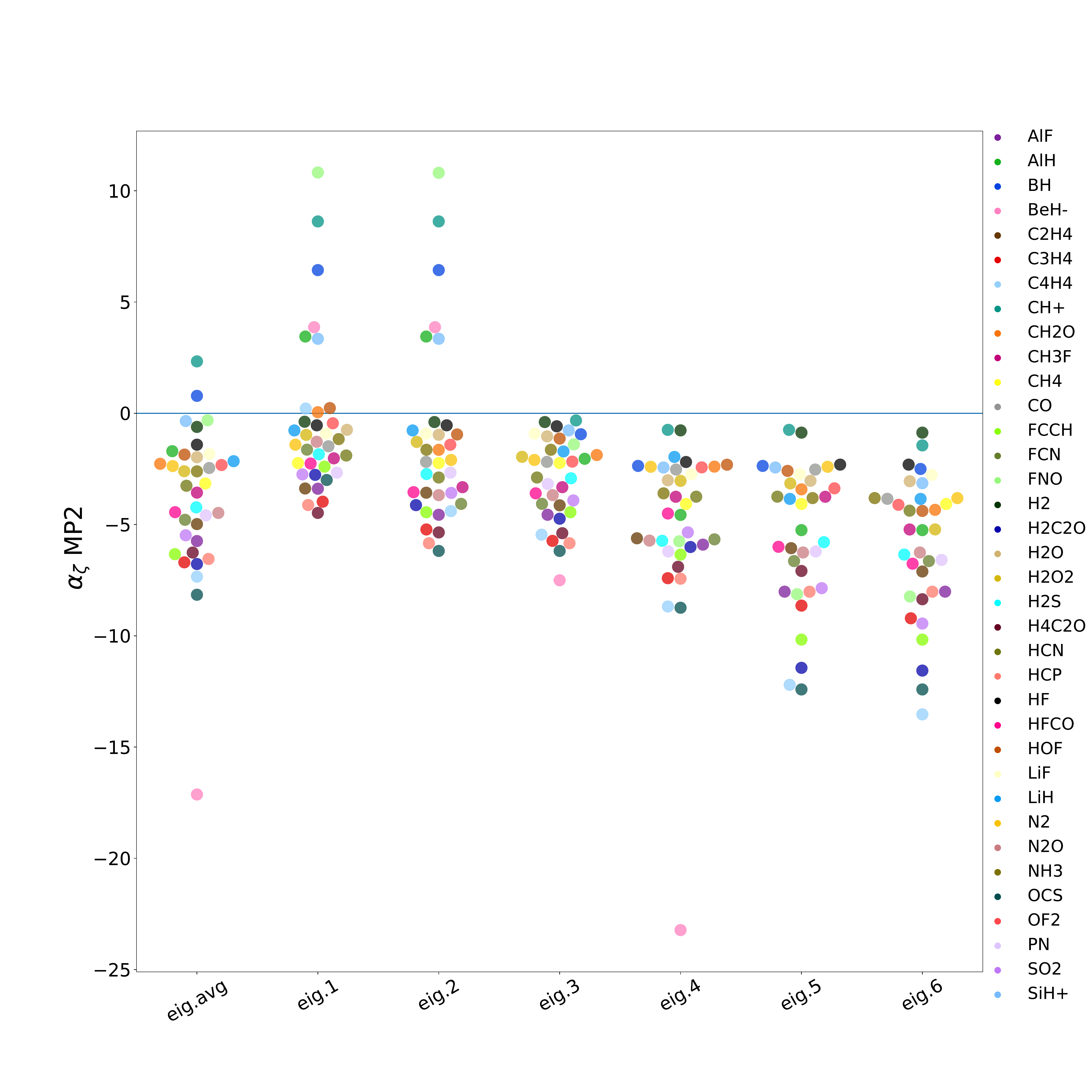}
    \caption{Eigenvalues of $\zeta$  in decreasing order computed with CCSD (top panel) and MP2 (bottom panel) in Luaug-cc-pCVTZ basis. eig avg is $\overline{\alpha}_\zeta$= $\frac{1}{6}$Tr($\zeta$). Eigenvalues no.\ 5 and 6 for BeH$^-$ lie beyond the range of the plot: eig5 = eig6 = -71.541 (top panel) and -79.810 (bottom panel).}
    \label{fig:eigAll_pl}
\end{figure}

Different response tensors in general have different physical dimensions and units, although this fact is somewhat obscured when working with atomic units. To account for this, we fix a length $\ell=a_0$, and define an auxiliary quantity $\mathbf{C}'=\ell \mathbf{C}$, and auxiliary response tensors $\overline{\mathcal{A}}=\ell^2 \mathcal{A}$ and $\overline{\mathcal{M}}=\ell \mathcal{M}$.
In atomic units, the numerical values of the quantities $\mathcal{A}$ and $\mathcal{M}$ remain unchanged by this transformation of response tensors to shared units.
Next, we construct a $6\times 6$ matrix of the form:
\begin{align}
    \boldsymbol{\zeta} = 
\begin{bmatrix}
\mathbf{\chi} & \mathbf{\overline{\mathcal{M}}} \\
\mathbf{\overline{\mathcal{M}}}^T & \frac{1}{2}\mathbf{\overline{\mathcal{A}}}
\end{bmatrix},
\end{align}
which allows us to reexpress the second-order energy in Eq.~\ref{eqEnergyExpansion} as
\begin{equation}
  \label{eqEnergyExpansionZeta}
  \begin{split}
   E(\mathbf{B},\mathbf{C}) & \approx E_0 + \frac{1}{2} \begin{bmatrix} \mathbf{B} \\ \mathbf{C}' \end{bmatrix}^T \begin{bmatrix} \mathbf{J}_{\mathbf{g}} \\ - \mathbf{a}/\ell \end{bmatrix} -  \frac{1}{2} \begin{bmatrix} \mathbf{B} \\ \mathbf{C}' \end{bmatrix}^T \boldsymbol{\zeta} \begin{bmatrix} \mathbf{B} \\ \mathbf{C}' \end{bmatrix}
  \end{split}
\end{equation}
The tensor $\boldsymbol{\zeta}$ is symmetric and has real eigenvalues, which we denote by $\boldsymbol{\alpha}_{\boldsymbol{\zeta}} = (\alpha_{\boldsymbol{\zeta};1},\ldots,\alpha_{\boldsymbol{\zeta};6})$.
\textit{If one or more eigenvalues are positive, i.e.\ energy decreases with any component of $(\mathbf{B},\mathbf{C})$, we classify the system broadly as paramagnetic. Otherwise, if all eigenvalues are negative, we classify the system as diamagnetic.}
Diagonalization of the submatrices gives us further details of this behaviour such as separate response to only $\mathbf{B}$ or $\mathbf{C}$.
Additionally, the trace of $\boldsymbol{\zeta}$ gives us a single number for the overall response to a generally non-uniform field.
We also obtain an average eigenvalue as
\begin{equation}
    \overline{\alpha}_{\boldsymbol{\zeta}} = \frac{1}{6} \sum_{i=1}^6 \alpha_{\boldsymbol{\zeta};i} = \frac{1}{6} \mathrm{Tr}(\boldsymbol{\zeta}),
\end{equation}
which is also the orientational average over all possible molecular orientations (with $\mathbf{B},\mathbf{C}$ fixed).
This procedure may be extended to increasingly non-uniform magnetic fields, such as those with curvatures and beyond.
In this paper, the magnetic field has a constant gradient and is of the form shown in Eq.~\ref{magfield}.

In Fig.~\ref{fig:eigAll_pl} we have plotted the eigenvalues, $\alpha_\zeta$, of the super-tensor, $\mathbf{\zeta}$ in decreasing order.
We have presented the values obtained with the most accurate methods we have studied in this paper, viz. CCSD and MP2. However, the classification is robust and all methods studied by us including Hartree-Fock (HF) and DFT show the same qualitative classification (except for FNO).
The geometry, energy and properties of FNO are all extremely sensitive to correlation and need at least CCSD(T) level computations for reasonable accuracy~\cite{Polak2008}. The values of the susceptibilities cannot be determined with any reasonable accuracy with our set of methods. Moreover, the computations do not converge with LDA and cM06-L functionals. We have thus left this molecule out of the error statistics presented in Sec.~\ref{results}.
All conventionally paramagnetic molecules show at least one positive eigenvalue.
In addition, FNO and HOF show positive eigenvalues arising from a paramagnetic response to some component of $\mathbf{C}$.
Thus, according to our criterion, the following molecules from our test set are paramagnetic: AlH, BH, BeH$^-$, C$_4$H$_4$, CH$^+$, CH$_2$O, SiH$^+$, HOF and FNO. The nature of this net paramagnetic behaviour is summarized in Table~\ref{tab:paraBC}.

We note that, due to the Cauchy interlace theorem, adding dimensions will increase the maximum eigenvalue and decrease the minimum eigenvalue. Hence, with $\alpha_{\boldsymbol{\zeta};\mathrm{max}} = \max_{1\leq i\leq 6} \alpha_{\boldsymbol{\zeta};i}$, $\alpha_{\boldsymbol{\chi};\mathrm{max}} = \max_{1\leq i\leq 3} \alpha_{\boldsymbol{\chi};i}$, and $\alpha_{\mathcal{A};\mathrm{max}} = \max_{1\leq i\leq 3} \alpha_{\mathcal{A};i}$, we have
\begin{equation}
     \alpha_{\boldsymbol{\zeta};\mathrm{max}} \geq \max \left\{\alpha_{\boldsymbol{\chi};\mathrm{max}},\ell^2 \frac{\alpha_{\mathcal{A};\mathrm{max}}}{2} \right\}.
\end{equation}
With similar notation for the minimum eigenvalues, we get an inequality in the reverse direction.
When $\mathcal{M}=\mathbf{0}$, usually by reason of symmetry, equality is achieved such as for C$_2$H$_4$, C$_4$H$_4$, CH$_4$, H$_2$ and N$_2$.
Equality is also achieved for the largest eigenvalue of $\mathbf{\zeta}$ in  CH$_2$O, CO and  HCP. 
The smallest eigenvalue also almost saturates in these cases.
This may indicate that there is a limit to how large the orbital paramagnetic and/or diamagnetic response can be within the limits of the basis set and molecular symmetry. While the non-zero components of $\mathcal{M}$ are small ($\sim 10^{-2}$~au) for CO and HCP, this is not the case for CH$_2$O ($\mathcal{M}_{xy}^{MP2}=0.956$, $\mathcal{M}_{yx}^{MP2}=0.760$~au).

Most molecules studied are highly diamagnetic with respect to $\mathbf{C}$ but we must remember that here we are only studying the orbital response.
The spin symmetry breaking caused by $\mathbf{C}$ will activate the spin-Zeeman term in the Hamiltonian leading to an overwhelmingly paramagnetic response to $\mathbf{C}$~\cite{Sen2018a}.
The interplay of spin and orbital effects of non-uniform magnetic fields is discussed in an earlier publication~\cite{Sen2018a}.
BeH$^-$ is particularly strongly diamagnetic to $\mathbf{C}$. The DFT computation with the cM06-L functional does not converge for BeH$^-$.
FNO, on the other hand, is strongly paramagnetic with respect to $\mathbf{C}$ according to computations with all the methods in our study.

\begin{table}
    \centering
    \begin{tabular}{|c|cc|cc|cc|cc|}
    \hline
         & \multicolumn{2}{|c|}{$\mathbf{B}$} & \multicolumn{2}{|c|}{$\mathbf{C}$} & \multicolumn{2}{|c|}{$(\mathbf{B},\mathbf{C})$}  \\ \cline{2-3} \cline{4-5} \cline{6-7}
     Molecule & $\overline{\alpha}_{\boldsymbol{\chi}}$ & $\alpha_{\boldsymbol{\chi};\mathrm{max}}$ & $\overline{\alpha}_{\mathcal{A}}$ & $\alpha_{\mathcal{A};\mathrm{max}}$ & $\overline{\alpha}_{\boldsymbol{\zeta}}$ & $\alpha_{\boldsymbol{\zeta};\mathrm{max}}$  \\ \hline
     AlH        & + & + & $-$ & $-$ & $-$ & +   \\
     BH         & + & + & $-$ & $-$ & + & +  \\
     BeH$^-$    & + & + & $-$ & $-$ & $-$ & +   \\
     C$_4$H$_4$ & + & + & $-$ & $-$ & $-$ & +   \\
     CH$^+$     & + & + & $-$ & $-$ & + & +   \\
     CH$_2$O    & + & + & $-$ & $-$ & $-$ & +   \\
     FNO        & $-$ & $-$ & + & + & $-$/+$^a$ & +   \\
     HOF        & $-$ & $-$ & $-$ & + & $-$ & +   \\
     SiH$^+$    & + & + & $-$ & $-$ & $-$ & +   \\
     \hline
    \end{tabular}
    \caption{Selected molecules which are proposed to be classified by us as paramagnetic. For response tensors related to $\mathbf{B}$, $\mathbf{C}$, and jointly to $(\mathbf{B},\mathbf{C})$, we show the sign of the average and maximum eigenvalue. $^a$The molecule FNO is extremely challenging for correlated theories, leading to different conclusions from different methods with respect to net response to $(\mathbf{B},\mathbf{C})$: diamagnetic ($-$) from CCSD,MP2, paramagnetic (+) from HF,KT3,cTPSS and not converged for LDA,cM06-L.}
    \label{tab:paraBC}
\end{table}

\subsection{Performance of Density Functional Approximations for Magnetic Susceptibilities}

In this section we discuss the relative performance of various wavefunction and density-functional methods in computing anapole magnetizabilities, $\mathcal{A}$ and $\mathcal{M}$. We compare this performance with their accuracy in describing the magnetizability, $\chi$ -- the most well-studied among the magnetic response quantities.

To quantify errors in the response tensors relative to a reference method we rely on the Frobenius norm,
\begin{equation}
 \epsilon_{\boldsymbol{\zeta}}^{\mathrm{method}} = \|\boldsymbol{\zeta}^{\mathrm{method}} - \boldsymbol{\zeta}^{\mathrm{ref}}\|_F = \left( \sum_{ij} |\zeta^{\mathrm{method}}_{ij} - \zeta^{\mathrm{ref}}_{ij}|^2 \right)^{1/2}
\end{equation}
and similarly for $\boldsymbol{\chi}$, $\mathcal{A}$, and $\mathcal{M}$. For the isotropic average, we use similar notation to mean $\epsilon_{\overline{\alpha}_{\boldsymbol{\zeta}}}^{\mathrm{method}} = |\overline{\alpha}_{\boldsymbol{\zeta}}^{\mathrm{method}} - \overline{\alpha}_{\boldsymbol{\zeta}}^{\mathrm{ref}}|$.

We present the error bars for the various methods as box and whisker plots where the median error is indicated by the horizontal line in the middle of the box and the top and bottom ends of the box indicate the third quartile and first quartile, respectively. The length of the box is thus the interquartile range. The top and bottom ends of the whiskers indicates the maximum and minimum errors considered in the estimation of the quartiles. Points beyond the whiskers are not considered in the statistics and are regarded as outliers.
We superimpose swarmplots on the box and whiskers plots to display the underlying data.

\begin{figure}
    \centering
    \includegraphics[width=\linewidth]{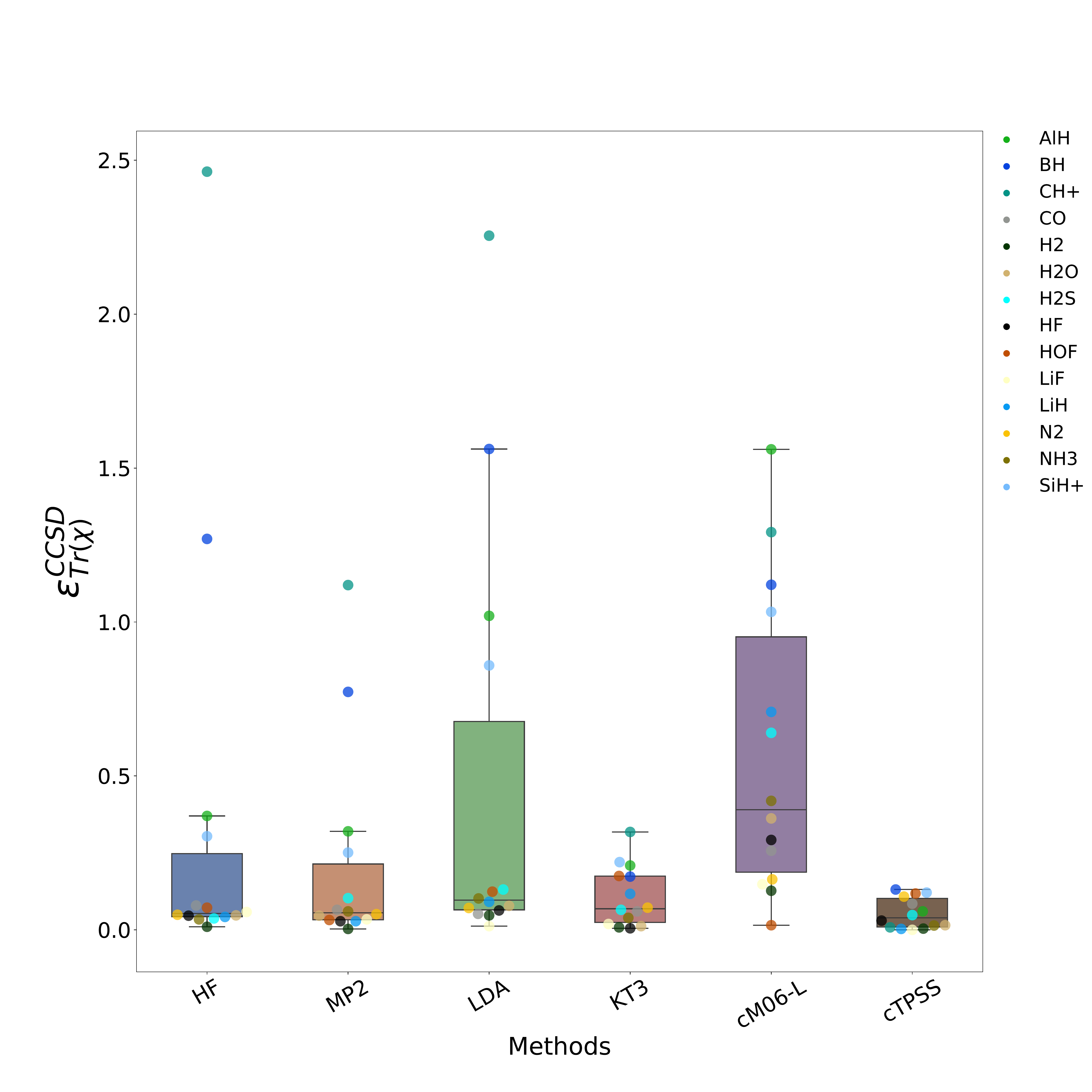}
    \includegraphics[width=\linewidth]{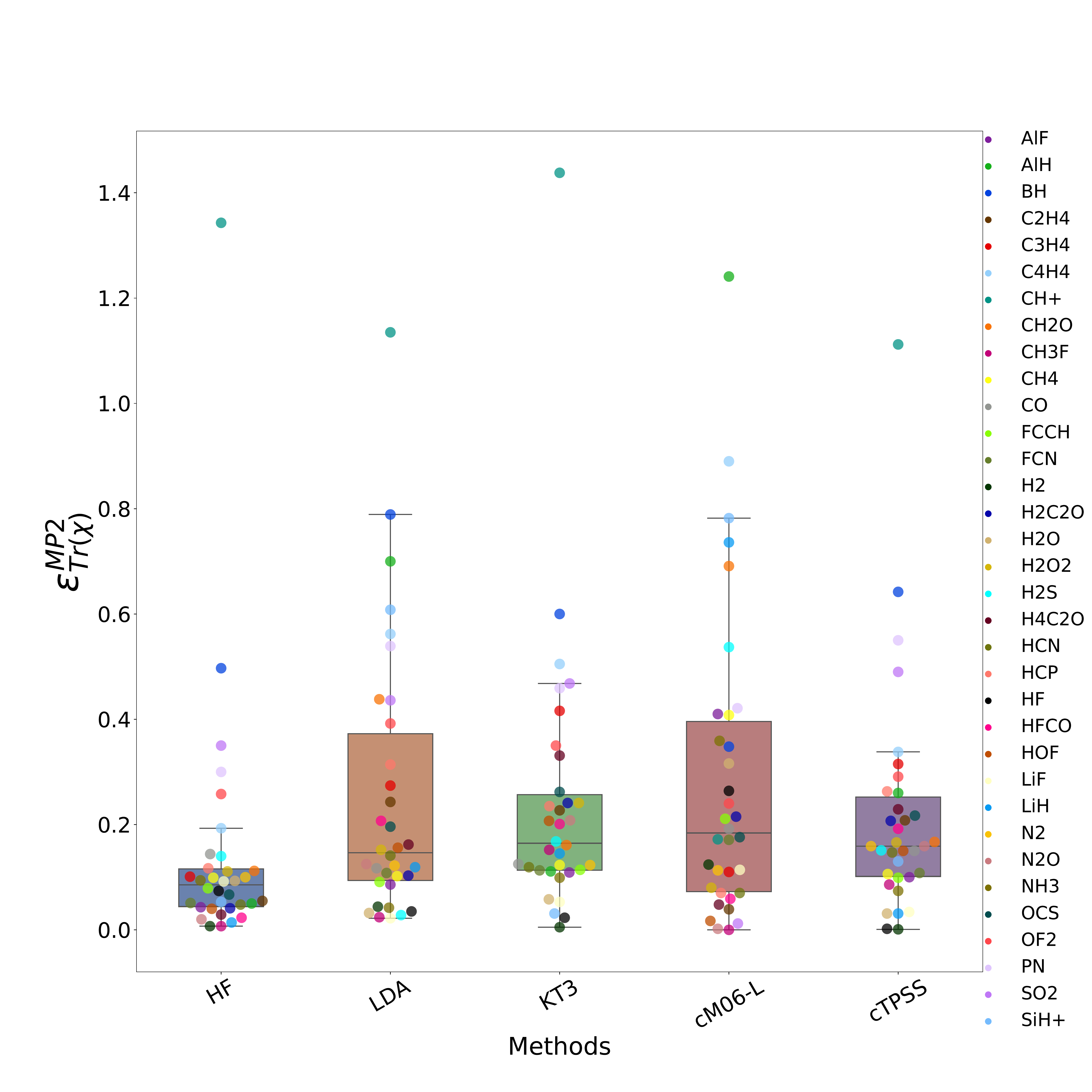}    
    \caption{Errors in the isotropic magnetizability computed by various methods in aug-cc-pCVTZ basis relative to CCSD (top panel) and MP2 (bottom panel).}
    \label{fig:isomagn_norm}
\end{figure}

Values of isotropic magnetizability computed with Hartree-Fock theory are often reasonable in the absence of low-lying excited states with correlation contributions of the order of 1-3\%~\cite{Gauss2007}.
However, DFT approximations which are reasonably good for correlation energy and electric properties, such as BLYP or B3LYP, are mostly inaccurate for magnetic properties, often being worse than Hartree--Fock theory.
This has prompted the development of exchange-correlation functionals tailored to magnetic properties such as the KT3 functional of Keal and Tozer~\cite{Keal2004}.
cTPSS has also been seen to be remarkably accurate for the same~\cite{Tellgren2014}.
Computational studies have indicated that the effect of electron correlation on the isotropic magnetizability ($\mathrm{Tr}(\mathbf{\chi}$)/3)~\cite{Gauss2007}.
is often an order of magnitude lower than that on the anisotropic magnetizability~\cite{Leszczynski}.
The isotropic magnetizability is also less sensitive to basis set size.
These conclusions are borne out by our results as shown in Fig.~\ref{fig:isomagn_norm}.
We have not studied BLYP and B3LYP as they are known to perform poorly for conventional magnetic properties.

\begin{figure}
    \centering
    \includegraphics[width=\linewidth]{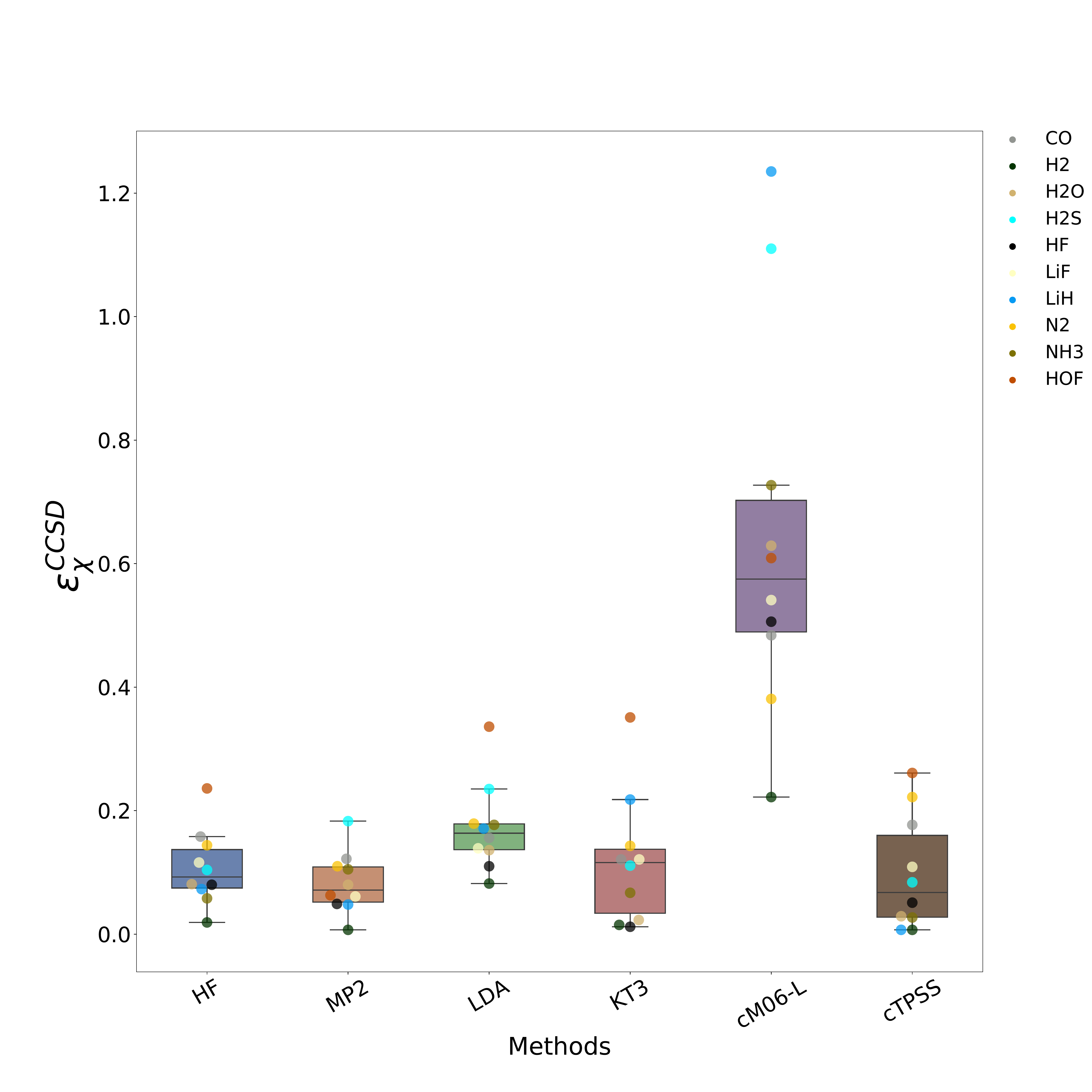}
    \includegraphics[width=\linewidth]{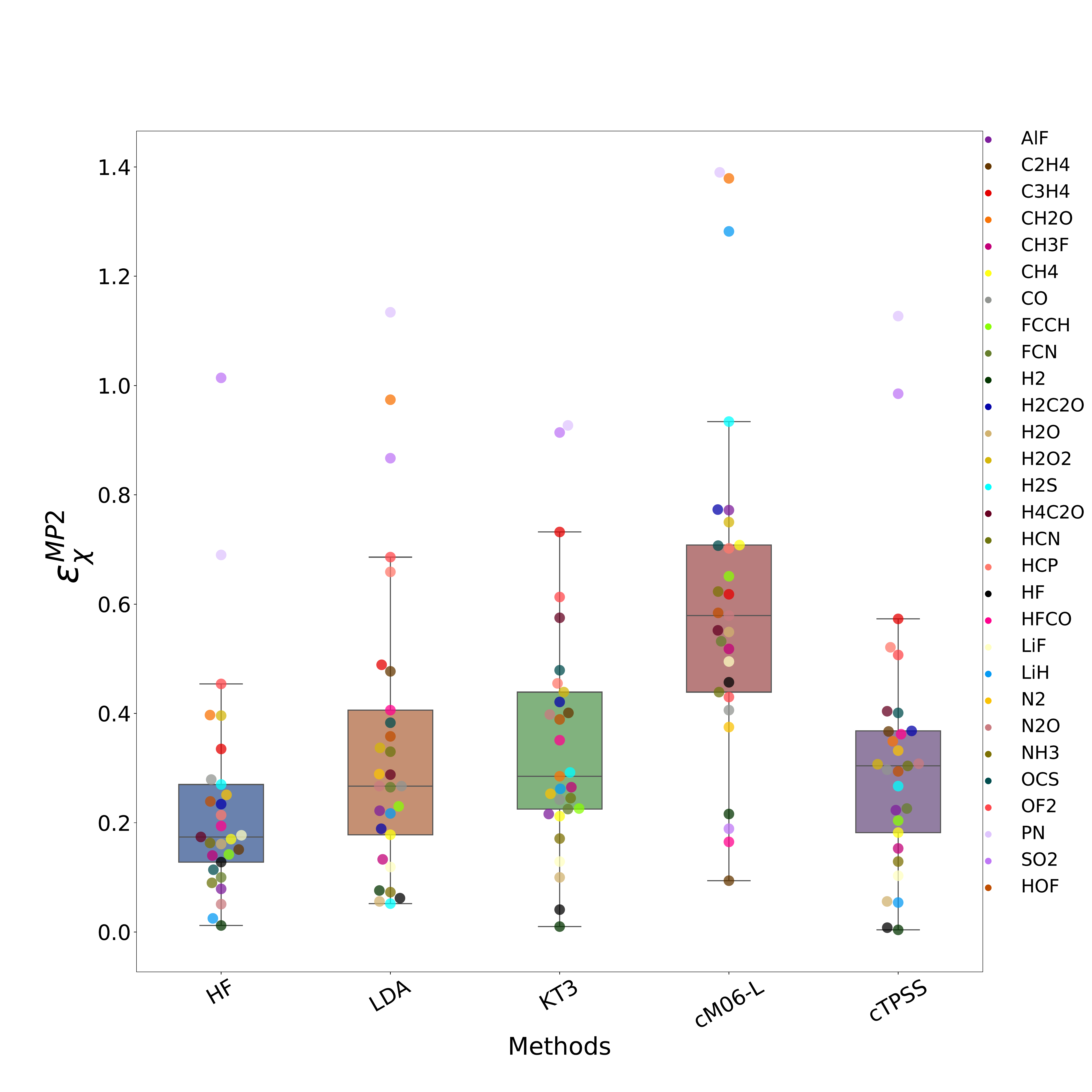}    
    \caption{Errors in magnetizability tensor, $\chi$, of the diamagnetic molecules in our test set computed by various methods in Luaug-cc-pCVTZ basis relative to CCSD (top panel) and MP2 (bottom panel).}
    \label{fig:dia_magn_norm}
\end{figure}

\begin{figure}
    \centering
    \includegraphics[width=\linewidth]{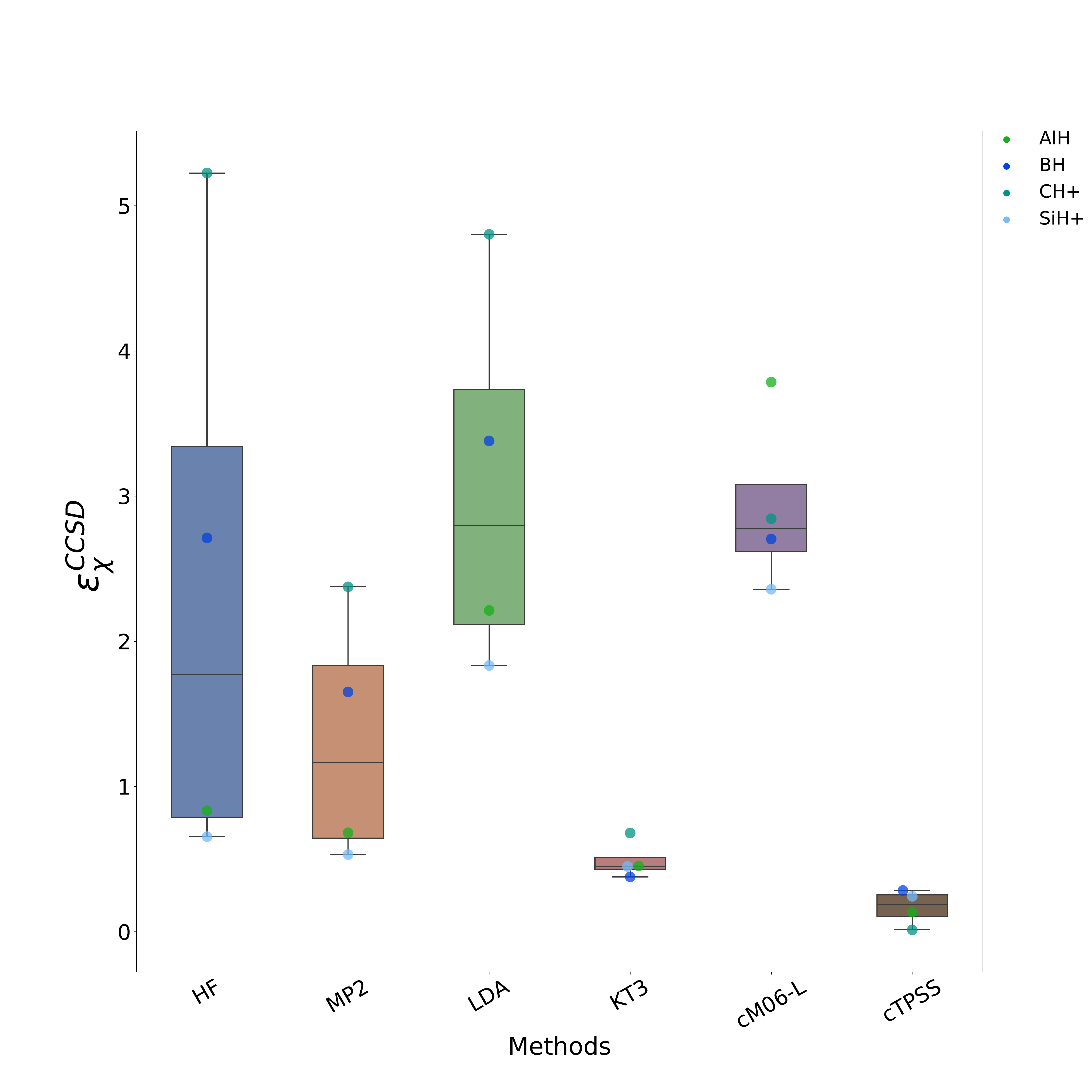}
    \includegraphics[width=\linewidth]{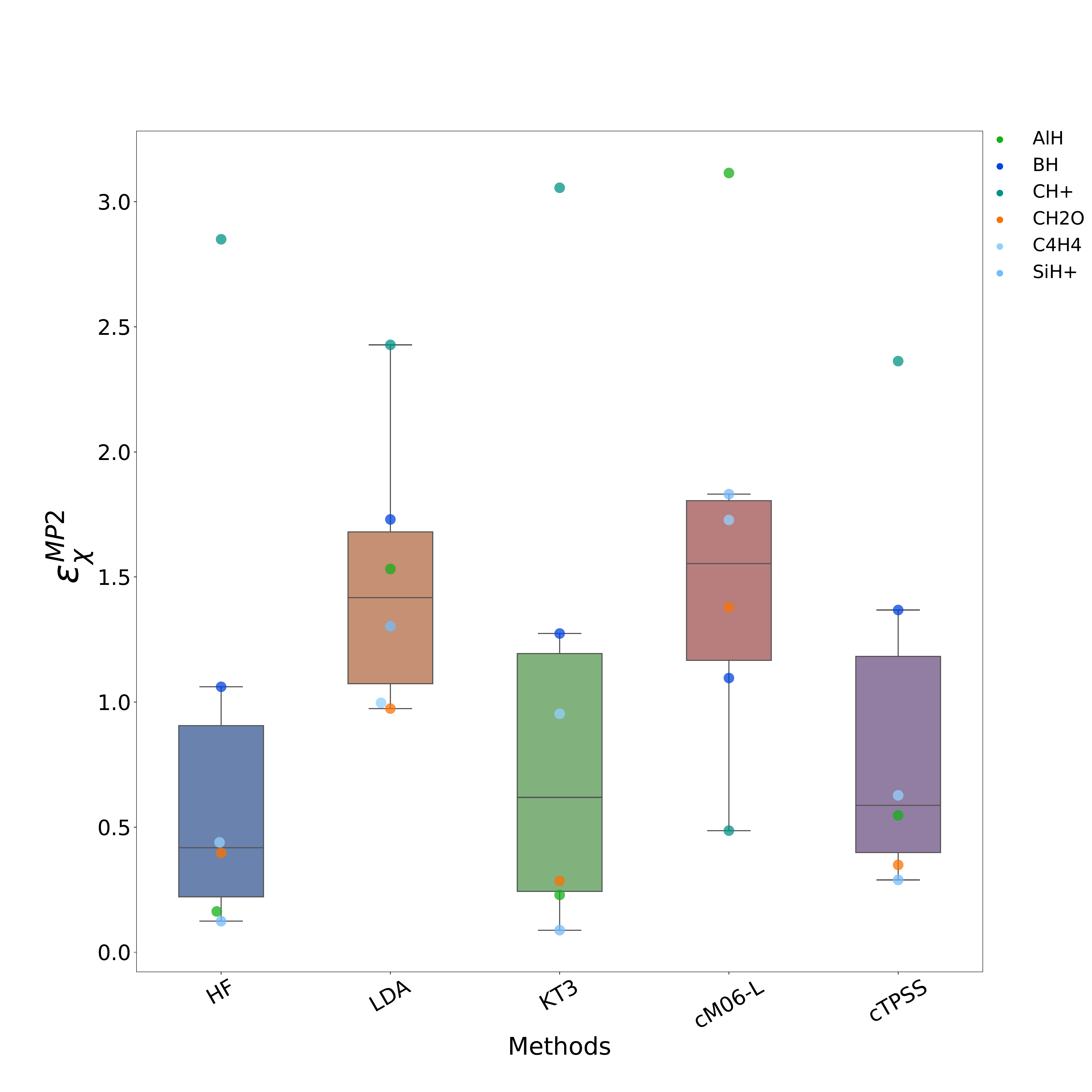}    
    \caption{Errors in magnetizability tensor, $\chi$, of the paramagnetic molecules in our test set computed by various methods in Luaug-cc-pCVTZ basis relative to CCSD (top panel) and MP2 (bottom panel).}
    \label{fig:para_magn_norm}
\end{figure}

While the accuracy of various DFT functionals for magnetizability has been well-studied, it remains to be seen if the same conclusions can be reached for more exotic properties such as anapole susceptibilities. In particular, we wish to explore if the KT3 and cTPSS functionals continue to perform well for $\mathcal{A}$ and $\mathcal{M}$.

In the top panel of Fig.~\ref{fig:dia_magn_norm} we can see that except the cM06-L functional all the methods considered here perform reasonably for the conventionally diamagnetic molecules. 
The cTPSS functional and MP2 show similar accuracy with HF and KT3 also doing quite well.
The bottom panel in Fig.~\ref{fig:dia_magn_norm} samples a larger test set using MP2 as the benchmark. Among the density functionals LDA, KT3 and cTPSS perform similarly.
It is interesting to note that the outlier for many of the methods in the top panel is HOF which would be classified as paramagnetic by our proposed scheme.
The trends for the paramagnetic molecules plotted separately in Fig.~\ref{fig:para_magn_norm} are much more surprising. All the errors are much higher than for the diamagnetic molecules indicating how much more difficult it is to describe the paramagnetic response.
MP2 no longer performs as well and is easily surpassed in accuracy by both KT3 and cTPSS. A similar conclusion was reached by Reimann et al.~\cite{Reimann2019} where even CCSD performed worse than cTPSS relative to CCSD(T) for the paramagnetic molecules. The bottom panel in Fig.~\ref{fig:para_magn_norm} is thus, not at all meaningful.
Between KT3 and cTPSS, cTPSS is somewhat more accurate for the paramagnetic systems studied by us.

\begin{figure}
    \centering
    \includegraphics[width=\linewidth]{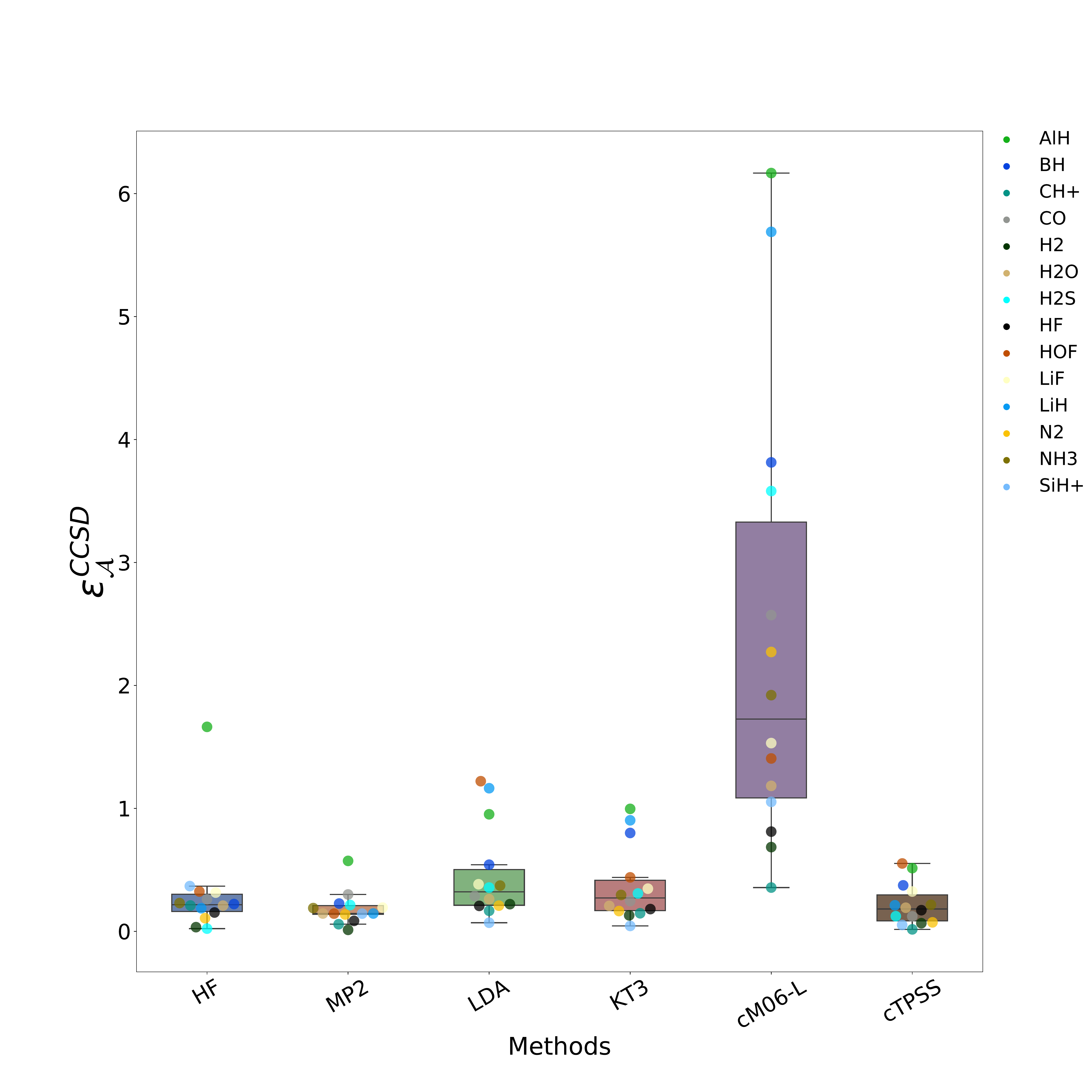}
    \includegraphics[width=\linewidth]{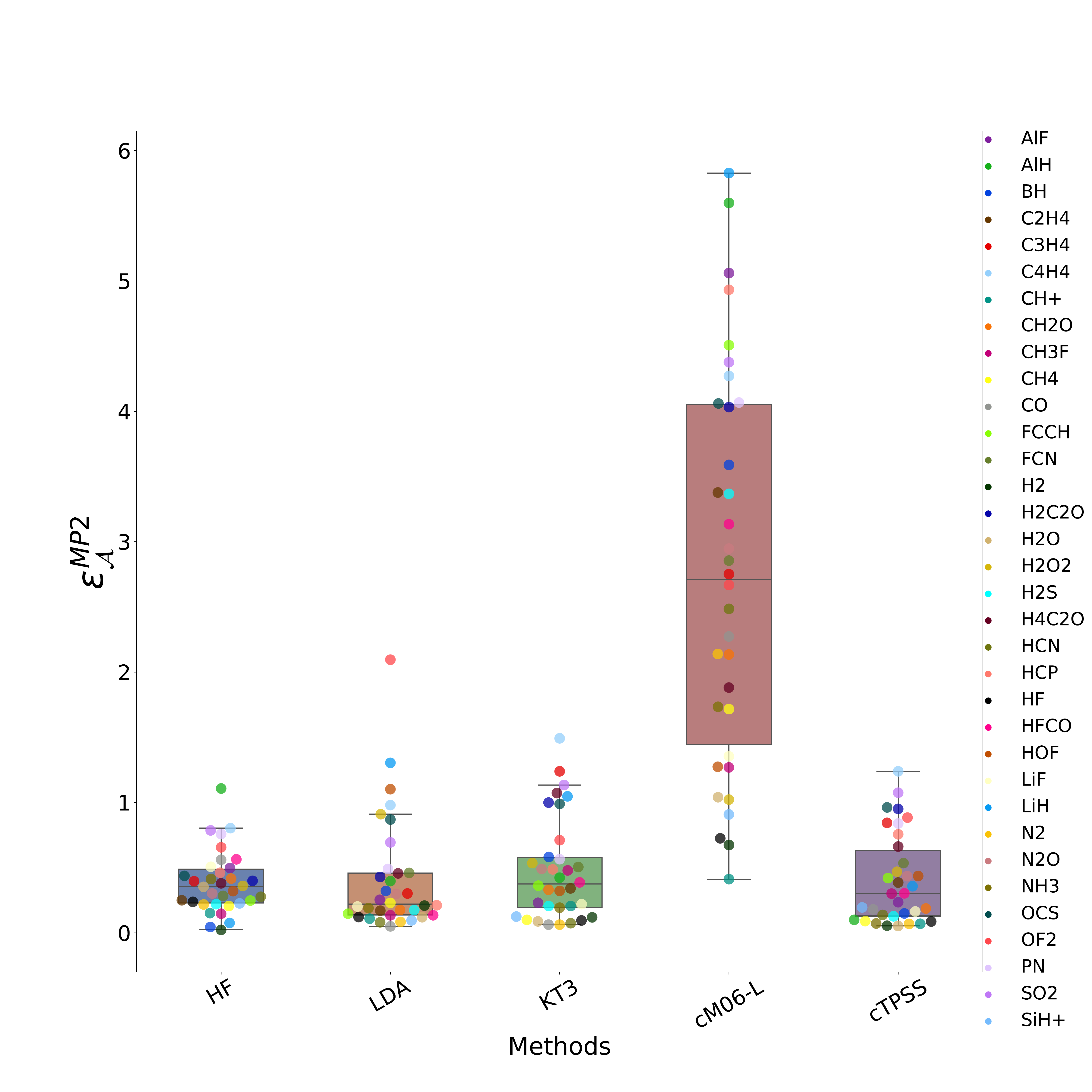}    
    \caption{Errors in anapole susceptibility, $\mathcal{A}$, computed by various methods in Luaug-cc-pCVTZ basis relative to CCSD (top panel) and MP2 (bottom panel).}
    \label{fig:anaA_norm}
\end{figure}

The top panel of Fig.~\ref{fig:anaA_norm} shows a reasonable description of $\mathcal{A}$ by most methods (except cM06-L) against CCSD. MP2 shows the highest accuracy with cTPSS following close behind. The bottom panel of Fig.~\ref{fig:anaA_norm} also follows the same trends with MP2 as reference.
Here too the paramagnetic molecules show a larger error than the diamagnetic ones with HOF behaving as the conventionally paramagnetic molecules.

\begin{figure}
    \centering
    \includegraphics[width=\linewidth]{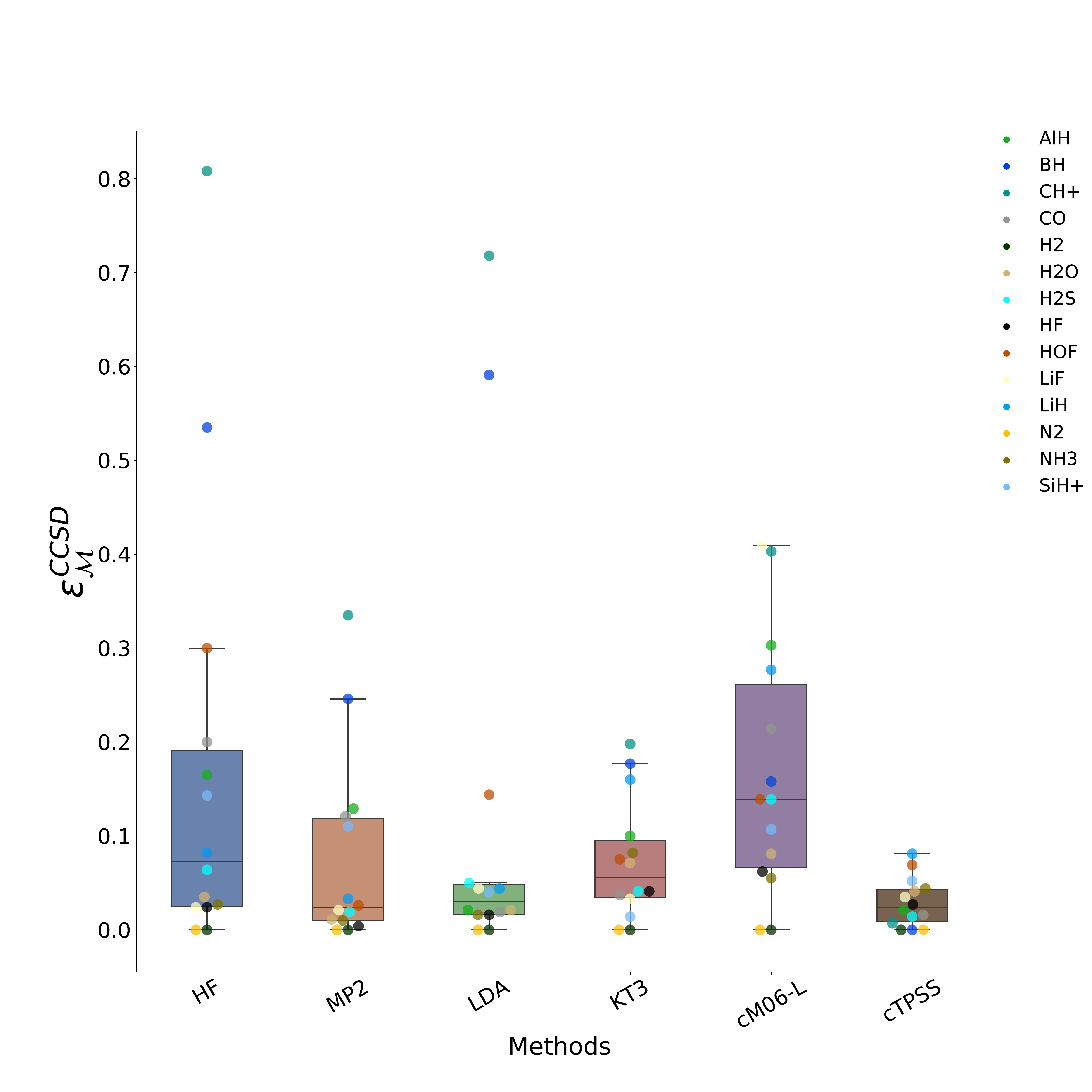}
    \includegraphics[width=\linewidth]{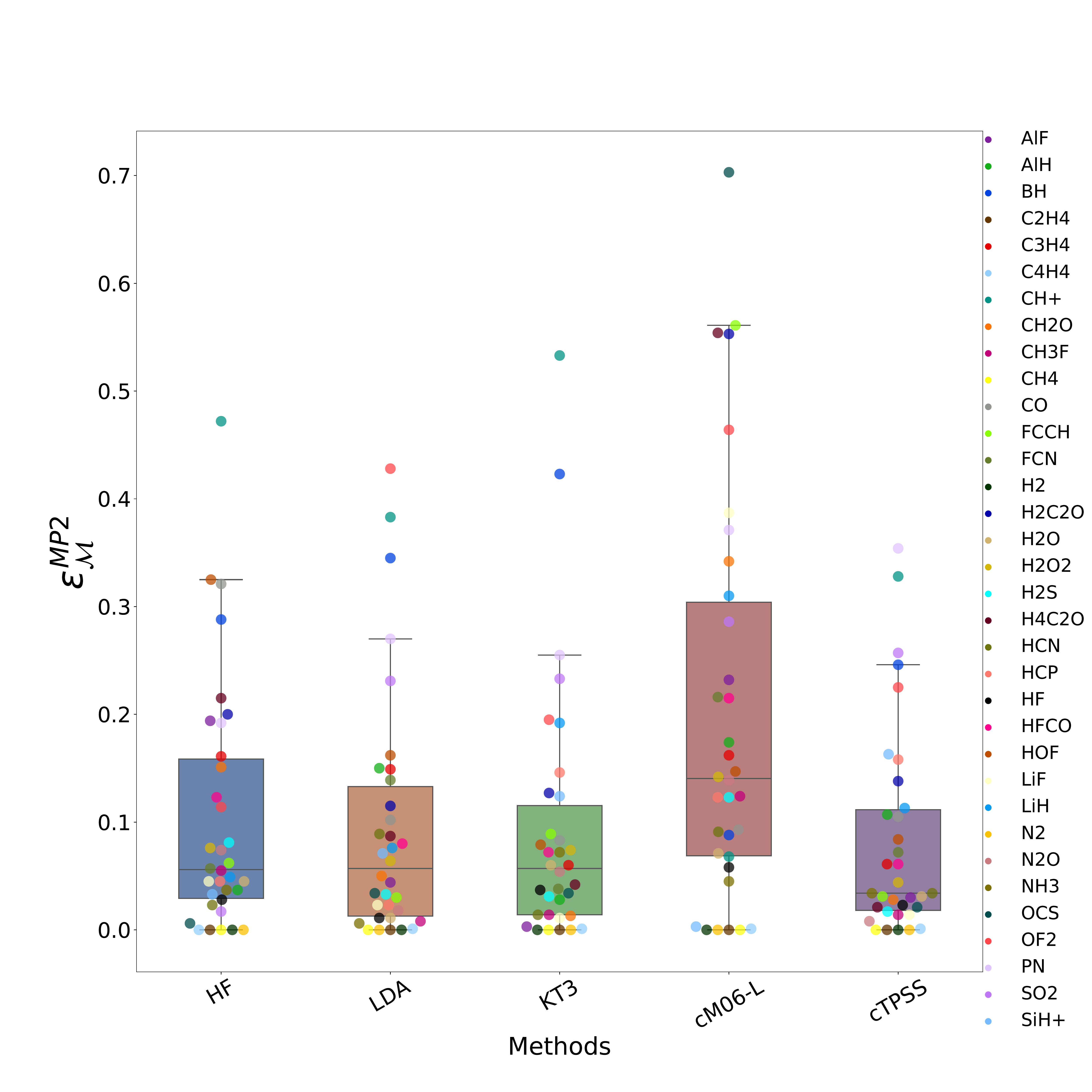}    
    \caption{Errors in mixed anapole susceptibility, $\mathcal{M}$, computed by various methods in Luaug-cc-pCVTZ basis relative to CCSD (top panel) and MP2 (bottom panel).}
    \label{fig:anaM_norm}
\end{figure}

Due to the smaller numerical values in the mixed anapole susceptibility tensor, $\mathcal{M}$, the errors appear to be smaller in Fig.~\ref{fig:anaM_norm}. cM06-L is no longer as bad although it is still the worst among the methods studied.

\begin{figure}
    \centering
    \includegraphics[width=\linewidth]{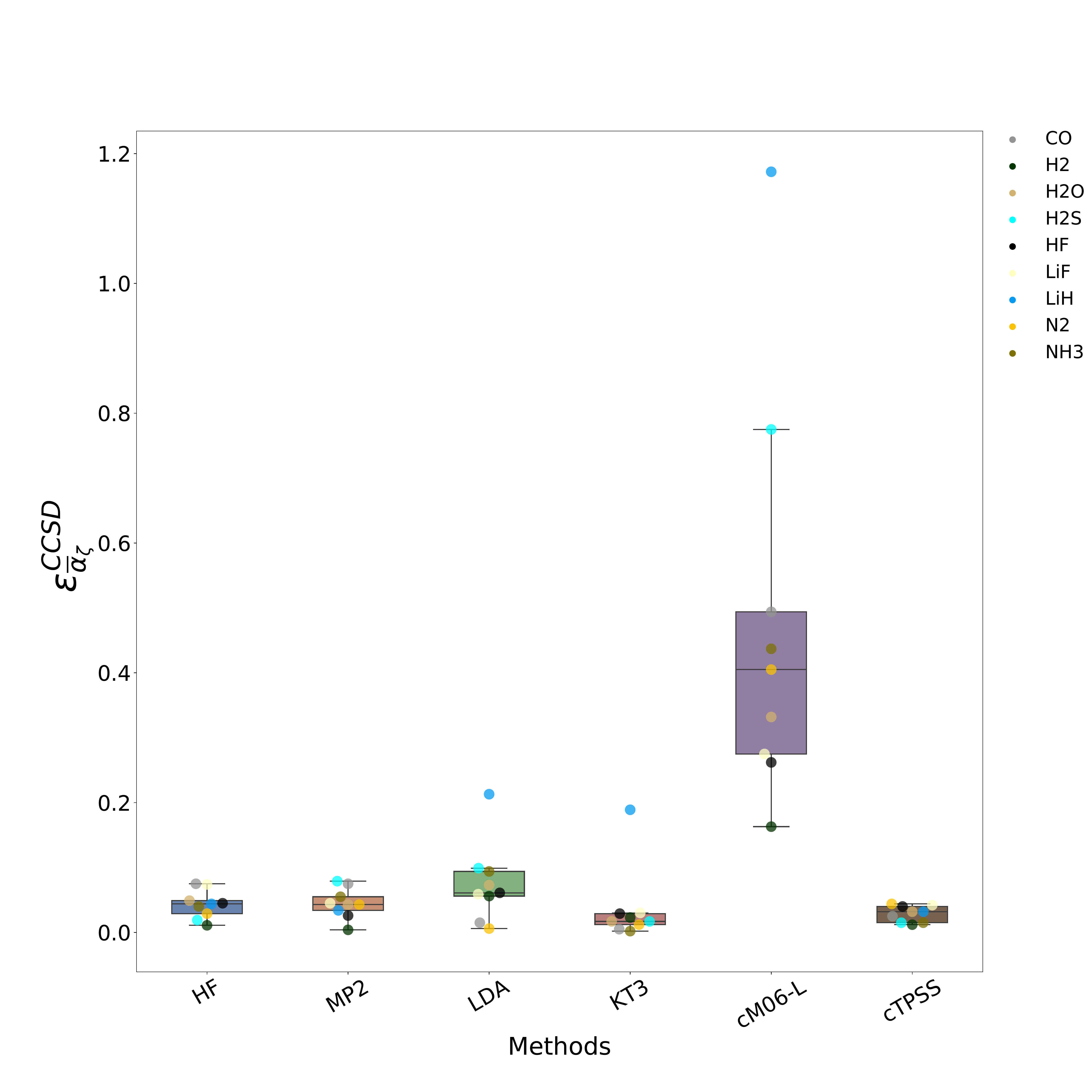}
    \includegraphics[width=\linewidth]{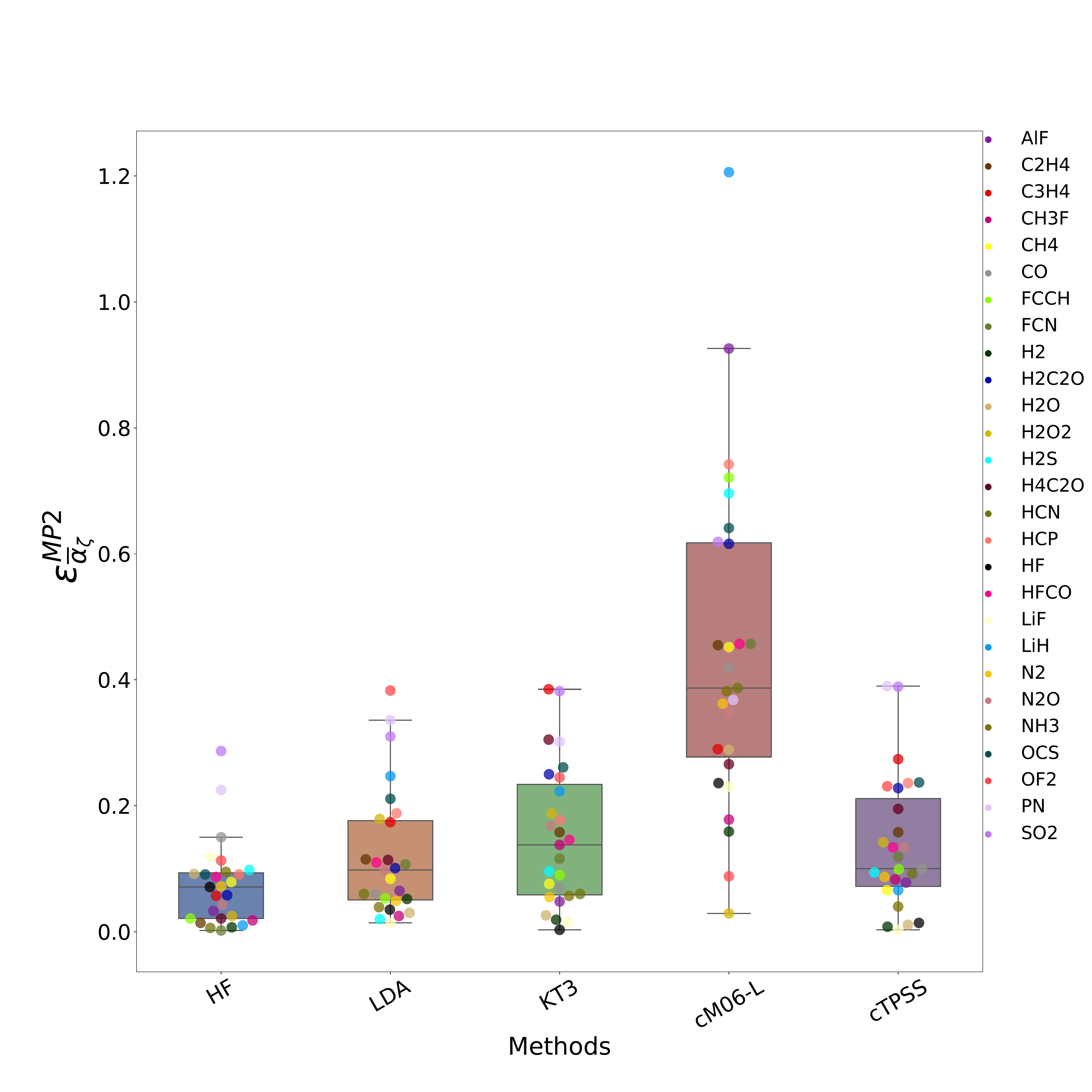}    
    \caption{Errors in the average eigenvalue of the super-tensor, $\zeta$, computed by various methods in Luaug-cc-pCVTZ basis relative to CCSD (top panel) and MP2 (bottom panel) for diamagnetic molecules.}
    \label{fig:dia_eigAvg_norm}
\end{figure}

\begin{figure}
    \centering
    \includegraphics[width=\linewidth]{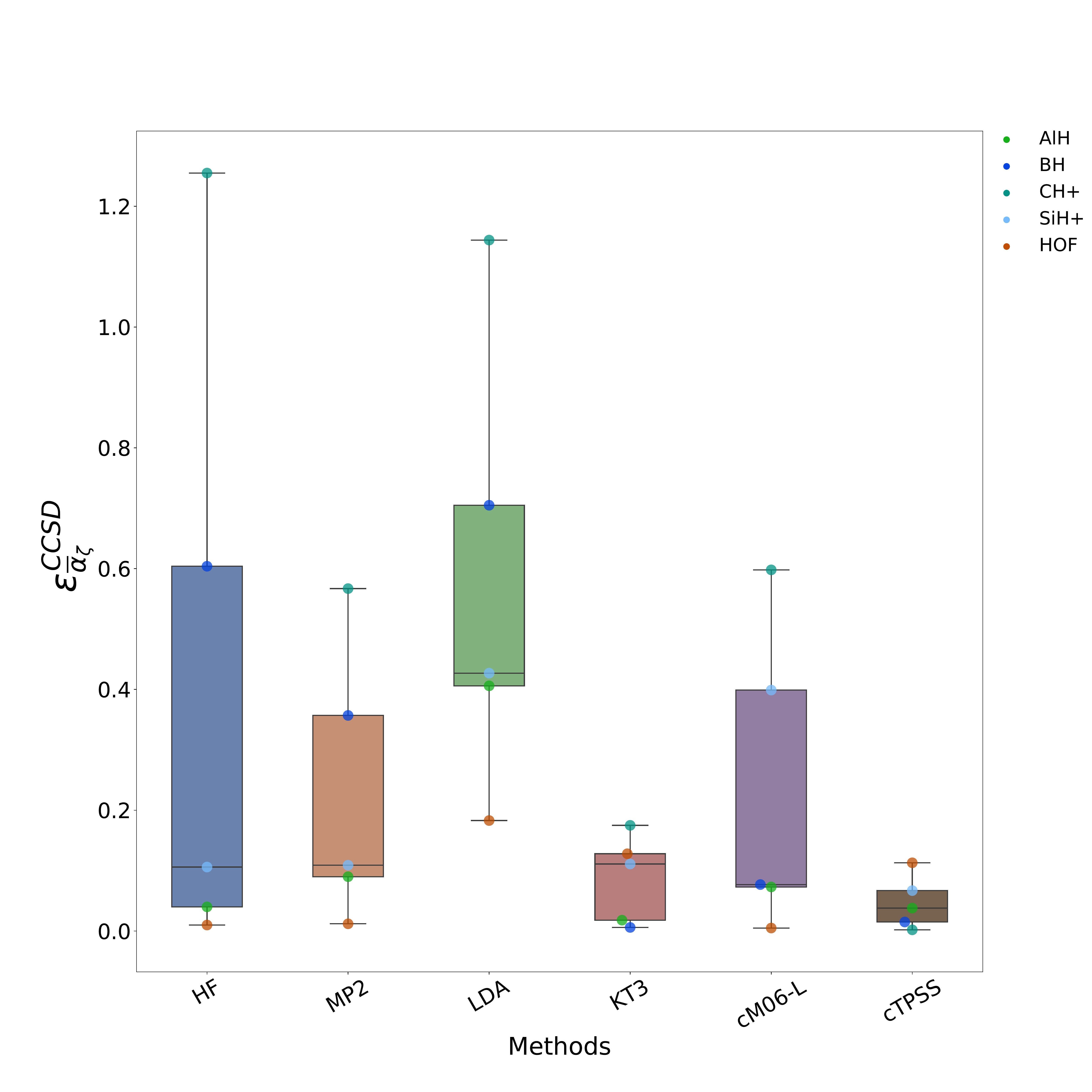}
    \includegraphics[width=\linewidth]{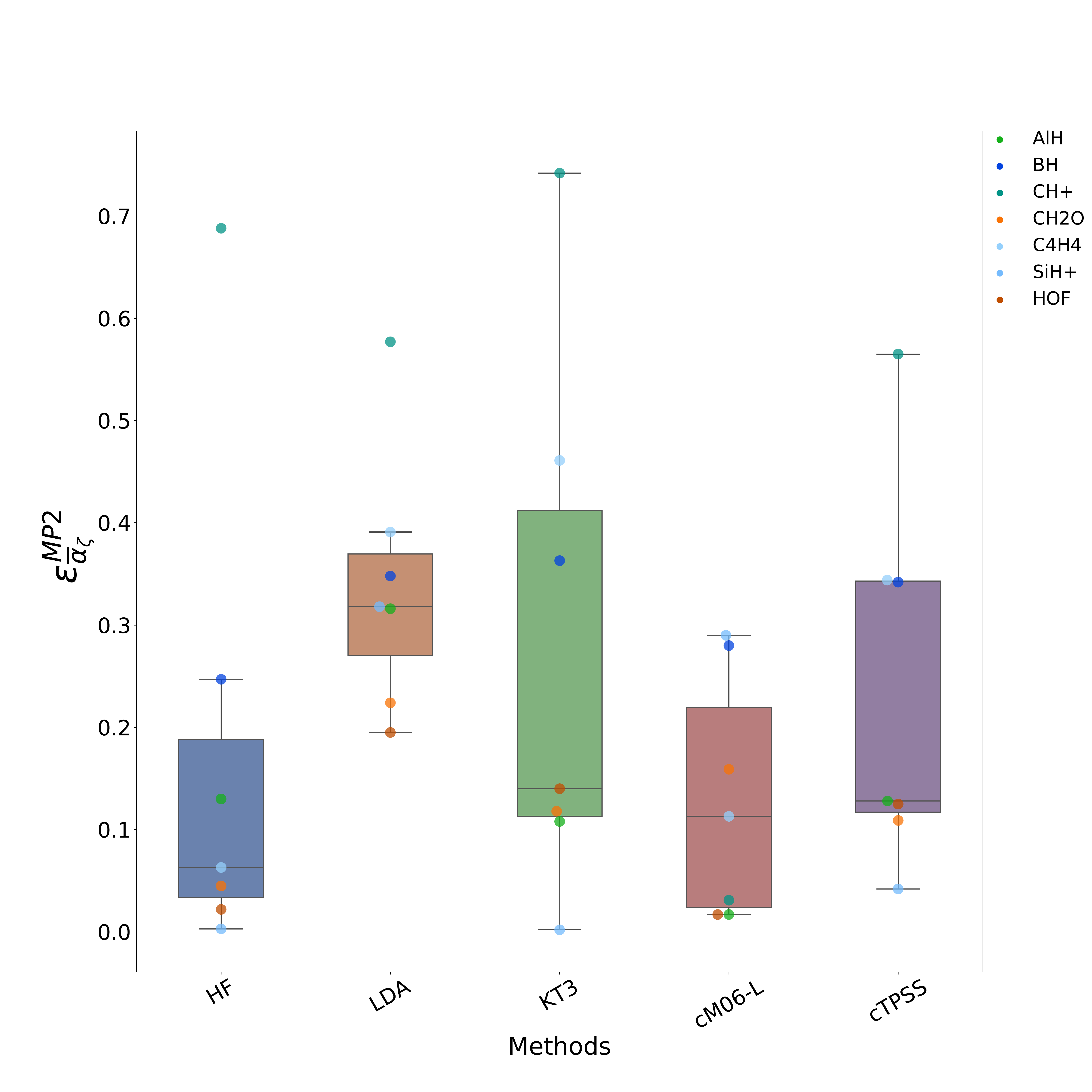}    
    \caption{Errors in the average eigenvalue of the super-tensor, $\zeta$, computed by various methods in Luaug-cc-pCVTZ basis relative to CCSD (top panel) and MP2 (bottom panel) for paramagnetic molecules.}
    \label{fig:para_eigAvg_norm}
\end{figure}

Finally, we present the errors in the quantity, $\overline{\alpha}_\zeta$ -- the average eigenvalue of our super-tensor, $\zeta$ in Figs.~\ref{fig:dia_eigAvg_norm} and \ref{fig:para_eigAvg_norm}.
The molecules have been classified according to the criterion in Sec.~\ref{classification}.
This may be considered as a condensed representation of all the errors presented in Figs.~\ref{fig:isomagn_norm}--\ref{fig:anaM_norm} allowing for the possibility of some error cancellation.
The top panel indicates a comparable performance of MP2, KT3
and cTPSS in comparison with CCSD. The larger test set in bottom panel also fits with a similar performance.
Molecules classified as paramagnetic by us again show the largest errors.

\subsection{Performance of Density Functional Approximations for Non-perturbative Effects}

\begin{figure}
    \centering
    \includegraphics[width=\linewidth]{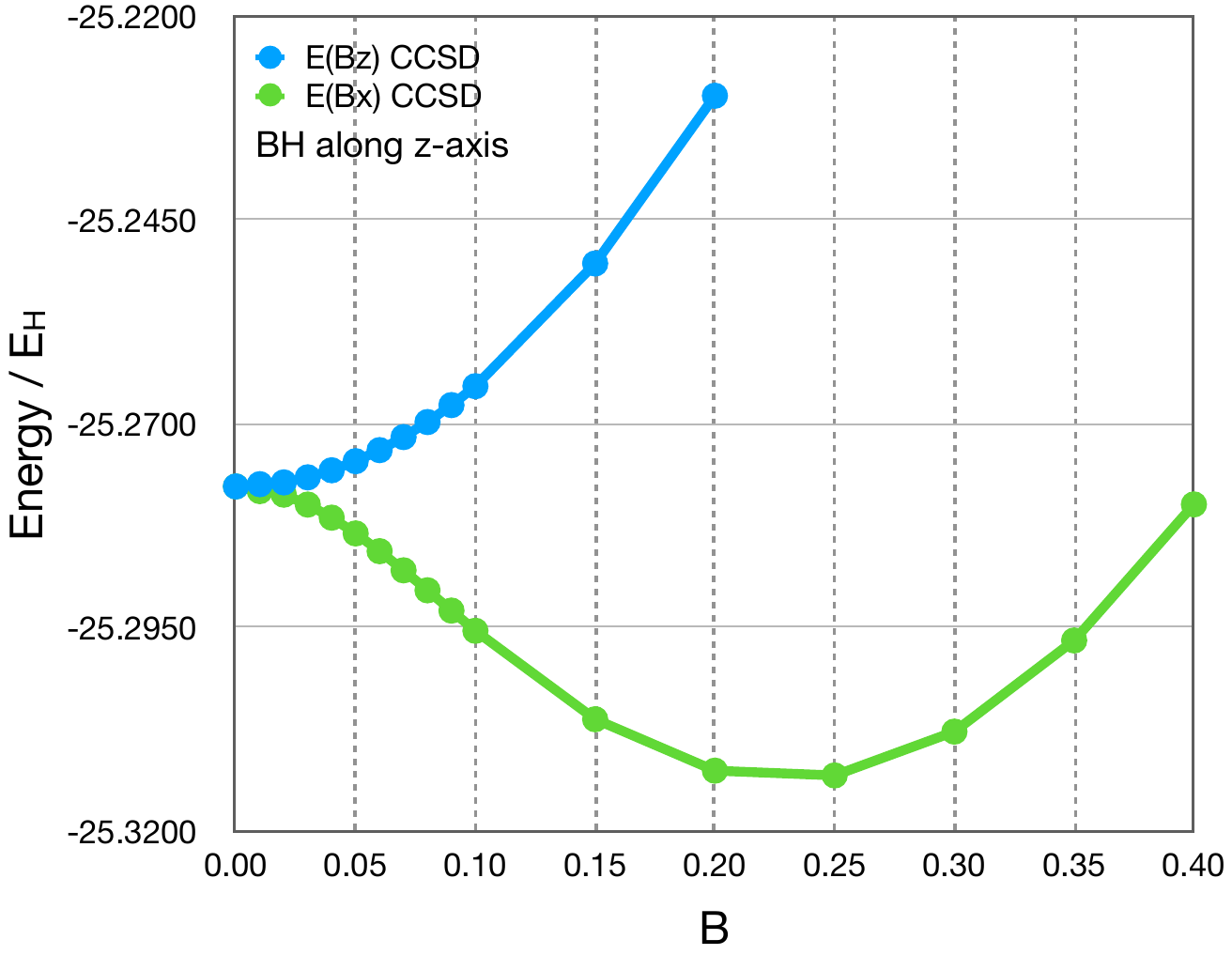}
    \includegraphics[width=\linewidth]{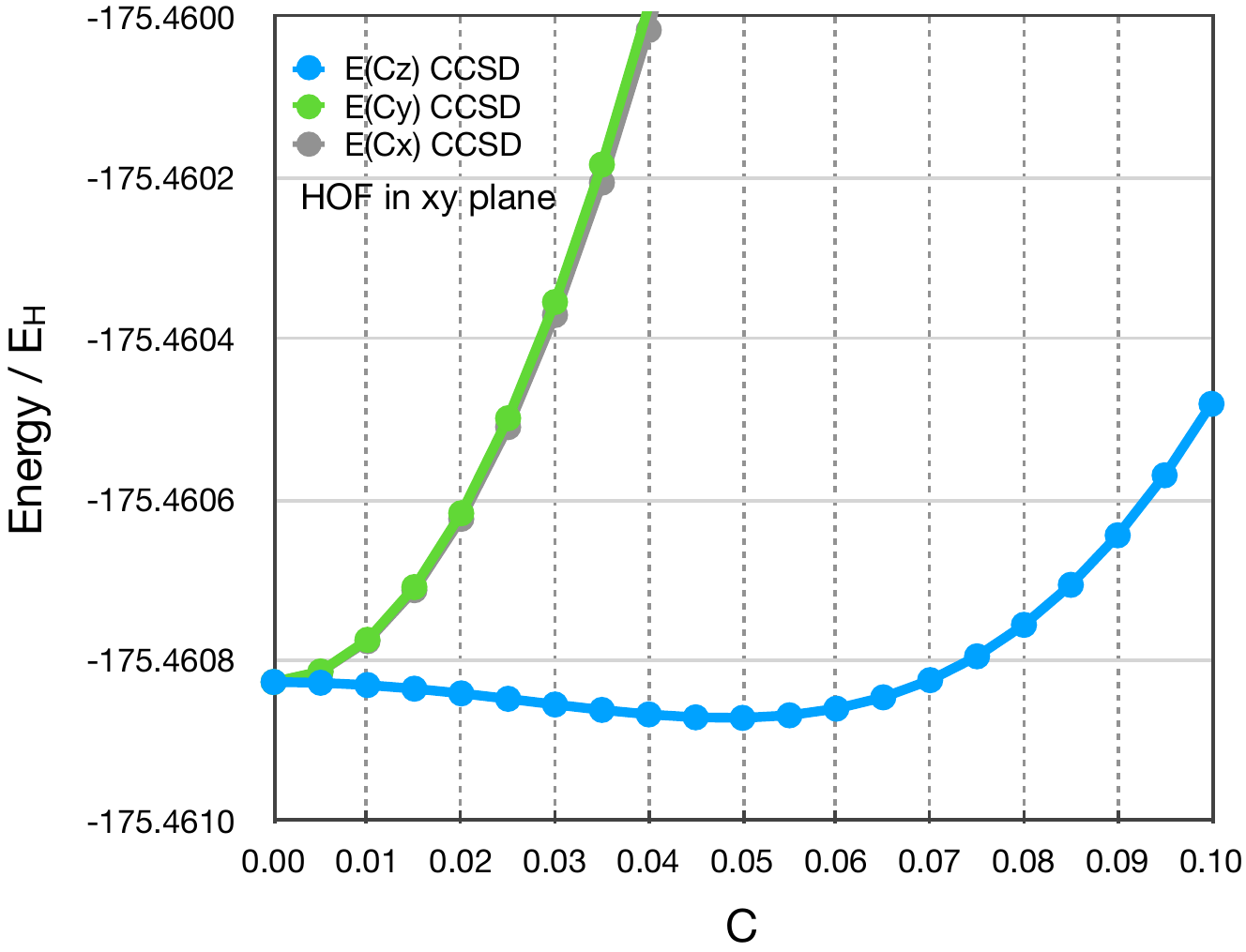}    
    \caption{Variation of energy of BH with a uniform field, \textbf{B} (top panel) and HOF with the curl of the field, \textbf{C} (bottom panel) showing initial paramagnetic orbital response and eventually the quadratic Zeeman effect in both cases.}
    \label{fig:energyplot}
\end{figure}

\begin{figure}
    \centering
    \includegraphics[width=\linewidth]{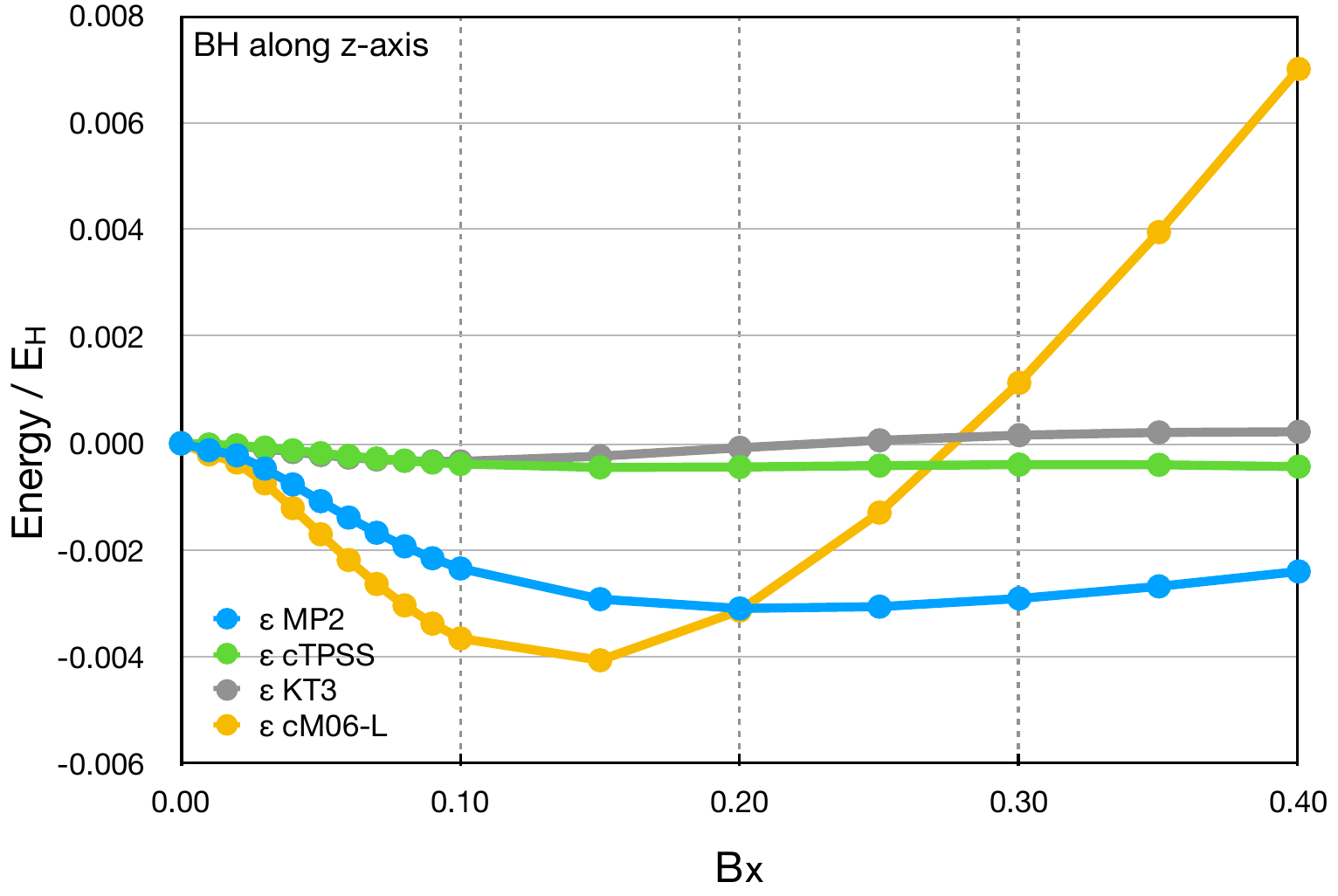}
    \includegraphics[width=\linewidth]{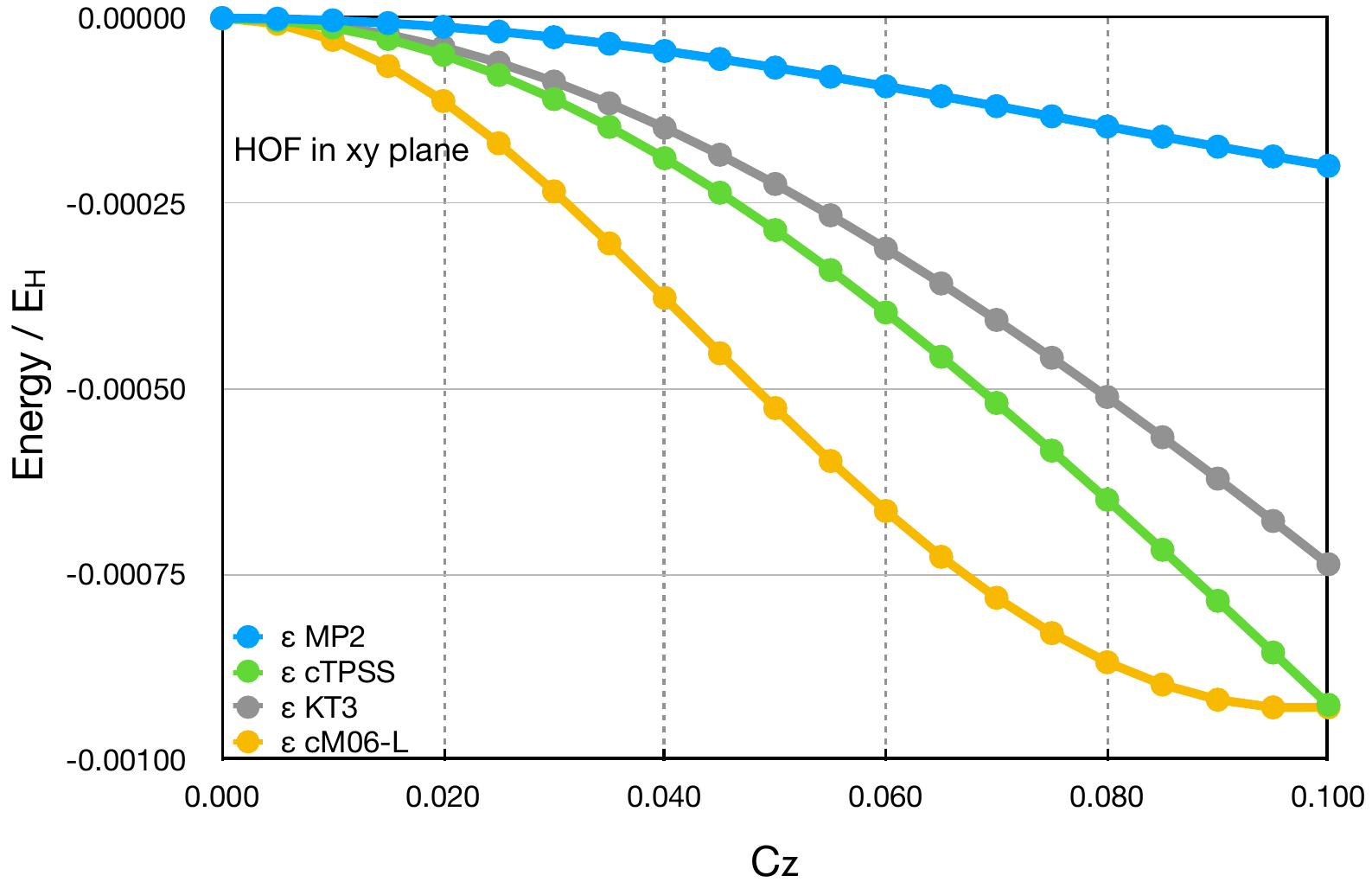}    
    \caption{Errors in energy, $\varepsilon$ relative to CCSD for BH with a uniform field, $B_x$ (top panel) and HOF with the curl of the field, $C_z$ (bottom panel). All energies have been shifted by the corresponding zero-field values such that all plots start at $\varepsilon=0$.}
    \label{fig:energyplotDFT}
\end{figure}

As we have seen in the previous subsection, molecules which show paramagnetic behaviour, with respect to a component of $\mathbf{B}$ or $\mathbf{C}$, are particularly challenging for all theories. In this subsection we further explore two molecules - BH as an example of paramagnetic behaviour with respect to a component of $\mathbf{B}$ and HOF as the newly discovered example of paramagnetic behaviour with respect to a component of $\mathbf{C}$.

In the top panel of Fig. \ref{fig:energyplot}, the paramagnetic behaviour of BH with respect to a field ($B_x$) perpendicular to the bond axis (z) is evident up to a critical field strength of about $B_x = 0.22$~au after which the quadratic term takes over. An analogous behaviour is observed in HOF (bottom panel of Fig. \ref{fig:energyplot}) with increase in the component of \textbf{C} ($C_z$) perpendicular to the molecular plane (xy). The turning point can be read off as $C_z = 0.048$~au. The depth of the minimum is, however, only $10^{-5}\ \mathrm{E}_H$ for HOF against a depth of $10^{-2} \ \mathrm{E}_H$ for BH.

We have tried to assess the capacity of various theories to describe the changes in electronic structure arising from the application of increasing $\mathbf{B}$ and $\mathbf{C}$. The CCSD method has been chosen as the reference and energies are computed with MP2 and a few selected density functional approximations. Since the zero-field energies themselves differ considerably we have subtracted the zero field energy computed with each method from all other data points thereby shifting all plots to a common starting point of zero. The energy difference between these shifted data points of various methods and CCSD is then plotted in Fig. \ref{fig:energyplotDFT}. For BH (top panel), cTPSS and KT3 work best with nearly parallel error curves. MP2 is surprisingly worse and the cM06-L functional gives a highly non-parallel error plot. For HOF (bottom panel), the error values themselves are an order of magnitude smaller than for BH with MP2 showing the best performance. KT3 and cTPSS follow the same trend of errors increasing with increasing $C_z$ as MP2. No flattening is observed even when HOF starts showing diamagnetic behaviour after the turning point of $C_z = 0.048$~au unlike the plots for BH. The cM06-L functional yields a very non-parallel error curve in this case too.
Although not shown here, the corresponding plots for BH vs $C_i, i=x,y,z$, also show increasing error with increasing $C_i$.

\section{Summary and Conclusion} \label{summary}

In this paper we have suggested a new classification of magnetic behaviour of molecules based on their response to a generally non-uniform field. We have demonstrated that paramagnetic behaviour can set in in a molecule due to inhomogeneities in the field even when its response to a uniform field is diamagnetic as is the case for FNO and HOF.
We have concluded that the susceptibilities of molecules---$\boldsymbol{\chi}$, $\mathcal{A}$ and $\mathcal{M}$---thus classified as paramagnetic are more difficult to describe.
Assuming that CCSD gives accurate results, KT3 and cTPSS are found to show the best performance among the DFT approximations with cTPSS being marginally better than KT3. The interquartile range for both functionals are narrow across all the properties studied by us.
cTPSS and KT3 also perform quite well for the more challenging paramagnetic molecules, even better than MP2 relative to CCSD.
cM06-L is particularly bad for magnetic properties performing even worse than LDA. These conclusions are also found to hold in the strong field regime as verified for some typical challenging molecules.
Hartree-Fock is surprisingly reliable for diamagnetic molecules with a more or less constant error across all magnetic properties.
The paramagnetic molecules are far more sensitive to correlation.
The eigenvalues of $\boldsymbol{\zeta}$, or even its higher dimensional analogues with additional parameters beyond $\mathbf{B}$ and $\mathbf{C}$, can serve as a concise measure for comparing accuracy of various theories in describing general magnetic properties.
Our findings are summarized in Table~\ref{tab:methoderrors}.

\begin{table}
    \centering
    \resizebox{\linewidth}{!}{%
    \begin{tabular}{|c|c|c|}
    \hline
      Property   & Class    & Error \\
    \hline
    $Tr(\chi)$  & dia   & cTPSS $<$ MP2 $<$ HF $\approx$ KT3 $<$ LDA $<<$ cM06-L \\
    $Tr(\chi)$  & para   & cTPSS $\approx$ KT3 $<$ MP2 $<$ HF $<$ LDA $\approx$ cM06-L\\
    $\chi$  & dia  & cTPSS $<$ MP2 $\approx$ HF $<$ KT3 $<$ LDA $<<$ cM06-L \\
    $\chi$  & para & cTPSS $<$ KT3 $<$ MP2 $<$ HF $<<$ LDA $\approx$ cM06-L \\
    $\mathcal{A}$  & dia   & MP2 $\approx$ cTPSS $<$ HF $<$ KT3 $<$ LDA $<<$ cM06-L  \\
    $\mathcal{A}$  & para   &  MP2 $<$ HF $\approx$ cTPSS $<$ KT3 $<$ LDA $<<$ cM06-L \\
    $\mathcal{M}$  & dia   & MP2 $<$ LDA $<$ cTPSS $<$ HF $<$ KT3 $<$ cM06-L \\
    $\mathcal{M}$  & para   &  cTPSS $<$ KT3 $<$ MP2 $\approx$ LDA $<$ HF $\approx$ cM06-L \\
    $\overline{\alpha}_\zeta$  & dia   & cTPSS $\approx$ KT3 $<$ MP2 $\approx$ HF $<$ LDA $<<$ cM06-L \\
    $\overline{\alpha}_\zeta$  & para  & cTPSS $\approx$ MP2 $\approx$ KT3 $<$ HF $<$ LDA $<$ cM06-L \\
    \hline
    \end{tabular}}
    \caption{Methods studied by us relative to CCSD in increasing order of median errors.}
    \label{tab:methoderrors}
\end{table}

\section*{Acknowledgements}

This work was supported by the Research Council of Norway through ``Molecular Spin Frustration'' Grant No. 240674, ``Magnetic
Chemistry'' Grant No. 287950, and CoE Hylleraas Centre for Molecular Sciences Grant No. 262695, and the European  Union’s Horizon 2020 research and innovation programme under the Marie Sk{\l}odowska-Curie grant agreement No.~745336. This work has also received support from the Norwegian Supercomputing Program (NOTUR) through a grant of computer time (Grant No.~NN4654K).

\end{document}